\documentclass[11pt]{article}
\usepackage{jheppub}
\usepackage{amsmath,amssymb,amsfonts,graphicx,slashed,color, mathtools, amsthm}
\usepackage{comment}
\usepackage{enumerate}
\usepackage{float}
\usepackage{caption}
\usepackage{subcaption}

\newcommand{\be}{\begin{equation}}
\newcommand{\ee}{\end{equation}}
\newcommand{\bea}{\begin{eqnarray}}
\newcommand{\eea}{\end{eqnarray}}
\newcommand{\beas}{\begin{eqnarray*}}
\newcommand{\eeas}{\end{eqnarray*}}
\newcommand{\ba}{\begin{array}}
\newcommand{\ea}{\end{array}}

\newcommand{\lads}{L_{\textnormal{AdS}}}

\title{Accelerating cosmology from $\Lambda <0$ gravitational effective field theory}

\author[1]{Stefano Antonini,}

\author[2]{Petar Simidzija,}

\author[3]{Brian Swingle,}

\author[2]{Mark Van Raamsdonk,}

\author[2]{Chris Waddell}

\affiliation[1]{Maryland Center for Fundamental Physics, University of Maryland, College Park, MD 20742, USA}
\affiliation[2]{Department of Physics and Astronomy, University of British Columbia,\\
6224 Agricultural Road, Vancouver, B.C.\ V6T 1Z1, Canada.}
 \affiliation[3]{Brandeis University, Waltham, MA 02453, USA}

\emailAdd{santonin@umd.edu}
\emailAdd{psimidzija@phas.ubc.ca}
\emailAdd{bswingle@brandeis.edu}
\emailAdd{mav@phas.ubc.ca}
\emailAdd{cwaddell@phas.ubc.ca}

\abstract{A large class of $\Lambda < 0$ cosmologies have big-bang / big crunch spacetimes with time-symmetric backgrounds and asymptotically AdS Euclidean continuations suggesting a possible holographic realization. We argue that these models generically have time-dependent scalar fields, and these can lead to realistic cosmologies at the level of the homogeneous background geometry, with an accelerating phase prior to the turnaround and crunch. We first demonstrate via explicit effective field theory examples that models with an asymptotically AdS Euclidean continuation can also exhibit a period of accelerated expansion without fine tuning. We then show that certain significantly more tuned examples can give predictions arbitrarily close to a $\Lambda$CDM model. Finally, we demonstrate via an explicit construction that the potentials of interest can arise from a superpotential, thus suggesting that these solutions may be compatible with an underlying supersymmetric theory.}

\begin{document}

\maketitle

\parskip=10pt

\section{Introduction}

In this paper, we consider the viability of models of cosmology based on gravitational effective theories associated to a dual CFT via the AdS/CFT correspondence. These EFTs have a negative extremum in their scalar potential (the negative cosmological constant, $\Lambda<0$) with an associated stable AdS solution. But they also have cosmological solutions of FRW form, including flat big bang matter/radiation cosmologies.
Naively, these wouldn't be realistic because of the negative cosmological constant. However, we will argue via examples that in the solutions that are most likely to have a holographic description, there will generically be time-dependent scalar fields whose potential energy can give rise to an accelerating phase in the cosmology without significant fine-tuning. We will show further that the resulting scale factor evolution can be realistic.

The initial motivation to study cosmological models based on EFTs with holographic duals is the goal of finding a complete quantum gravity for {\it some} four-dimensional big bang cosmology, realistic or not. Since some solutions of these $\Lambda < 0$ effective theories have a known microscopic description via holography, it is plausible that holographic tools may play a role in coming up with a microscopic description of the cosmological solutions as well.\footnote{On the other hand, it is presently much less clear how models with a positive cosmological constant can be given a microscopic description; see \cite{Banks:2001px,Strominger:2001pn,Alishahiha:2004md,Gorbenko:2018oov,Coleman:2021nor,Freivogel2005,McFadden:2009fg,Banerjee:2018qey,Susskind:2021dfc} for various approaches.} 

There is a strong hint for how holography might play a role: cosmological solutions with $\Lambda < 0$ and some combination of matter and radiation are time-symmetric big-bang / big-crunch spacetimes whose analytic continuation are real geometries with asymptotically AdS regions for Euclidean time $\tau \to \pm \infty$; following \cite{Maldacena:2004rf} we will refer to these geometries as wormholes (see \cite{McInnes:2004nx} for an early study of accelerating cosmologies in similar setups). This suggests a picture where the Euclidean theory is defined holographically via a pair of 3D CFTs and observables in the cosmology are related by analytic continuation to observables in this Euclidean theory. We have suggested a possible framework for this description in \cite{Cooper:2018cmb, Antonini2019,VanRaamsdonk:2021qgv, Antonini:2022blk, Antonini:2022xzo}, but we will not assume this in the present paper.\footnote{Possible holographic constructions must address the factorization puzzle \cite{Maldacena:2004rf}; suggested resolutions include considering an ensemble of CFTs or a direct interaction between the CFTs \cite{Maldacena:2004rf,Betzios:2019rds,VanRaamsdonk:2020tlr}.}

The existence of dual CFTs associated to the asymptotically AdS regions in the Euclidean solution suggests that the solutions will often also have non-trivial scalar fields \cite{Antonini:2022blk,VanRaamsdonk:2022rts}. Scalars with $m^2 < 0$ will be present in the effective field theory in the fairly generic situation that the CFTs have relevant scalar operators.\footnote{Recall that the scalar masses are related to operator dimensions by $m^2 \lads^2 = \Delta (\Delta - 3)$.} These scalar fields vanish at the asymptotically AdS boundary, but in the most generic asymptotically AdS solutions, the $m^2 < 0$ scalars turn on as we move away from the boundary. This is associated with RG flow in the dual CFT. The radial dependence of the scalar field in the wormhole translates to a time-dependence of the scalar field in the cosmology.

The main goal of this paper is to show that these scalar fields can naturally give rise to an accelerating period in the cosmology, and that in some cases, the acceleration can be realistic, with the cosmological scale factor matching arbitrarily well with the scale factor of conventional models that provide a good fit to observational data.\footnote{See \cite{Cardenas:2002np} for an early study of accelerating negative $\Lambda$ cosmologies compatible with data.}

\begin{figure}
    \centering
    \includegraphics[width=1\textwidth]{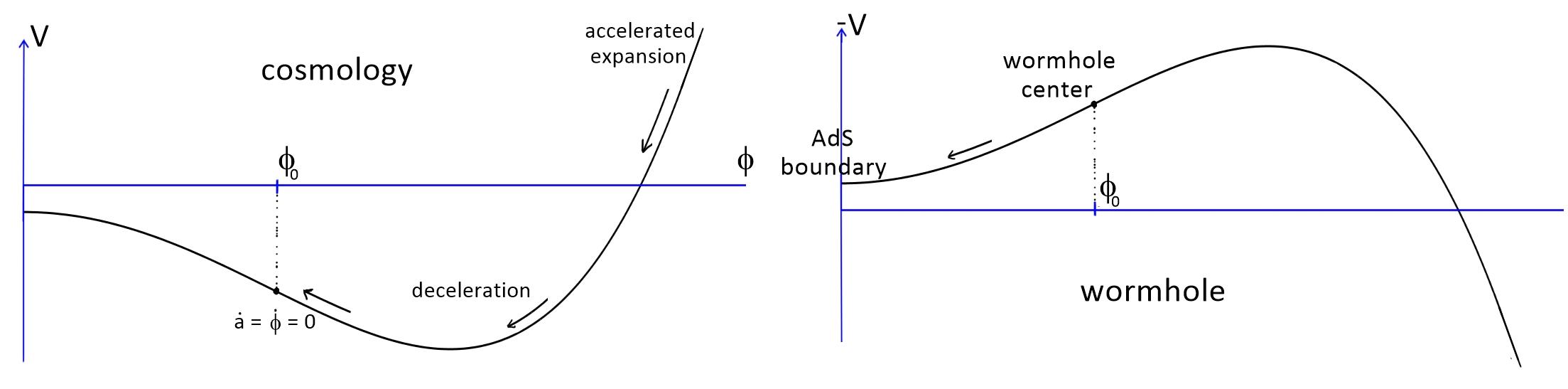}
    \caption{Evolution of the scalar field in a typical potential in the cosmology and its Euclidean continuation. }
    \label{fig:scalar}
\end{figure}

A typical potential for a scalar field associated with a relevant CFT operator is shown in Figure \ref{fig:scalar} (left). The negative $m^2$ gives the quadratic decrease in $V$ for small $\phi$, while the eventual increase is due to the interaction terms that would typically be present; for example, a higher-order polynomial potential should eventually be positive and growing if the EFT is stable.

In the Euclidean solution, the evolution of the scalar field from the middle of the wormhole to the AdS boundary corresponds to the damped motion of a particle in the potential $-V$ with damping constant $3 a' / a \equiv 3 (da/d\tau)/a$. In order for this extremum to be associated with a stable AdS solution, we must satisfy the Breitenlohner-Freedmann bound $V'' > -9/(4L_{AdS}^2)$, which turns out to correspond to the fact that the motion of the scalar field towards the extremum is overdamped. Thus, the simplest evolution of the scalar in the wormhole is a simple monotonic evolution from a lower value of the potential (i.e. a higher value of $-V$) at the middle to the higher extremal value at the boundaries, as shown in Figure \ref{fig:scalar} (right).

In the cosmology, the value of the scalar at the time-symmetric point is the same as the value of the scalar at the middle of the Euclidean wormhole. The evolution back towards the big bang or forwards towards the big crunch corresponds to the anti-damped motion of the scalar in the potential $V$, initially moving in the direction towards the positive part of the potential. In many cases, the scalar will reach the positive region of the potential before the scale factor goes to zero, so 
part of the cosmological evolution has positive dark energy. 

This opens up the possibility that some of these $\Lambda < 0$ models might exhibit an accelerating phase (as we have argued in \cite{Antonini:2022blk, Antonini:2022xzo}). To check this, we study various explicit examples in Section \ref{sec:realistic}. We first verify that solutions based on the scalar evolution shown in Figure \ref{fig:scalar} can lead to periods of accelerated expansion in the cosmology without excessive fine-tuning. We find examples of simple potentials that give rise to acceleration for generic values of their parameters, meaning in a codimension zero region of the parameter space.

Next, we check that by taking an appropriately chosen potential, we can find models with acceleration where the scale factor is arbitrarily close to $\Lambda$CDM or $w$CDM models for the past evolution. These models are finely tuned, but we include them as a demonstration that the framework can in principle give results consistent with observations. For realistic cosmology, we want to match observations, not $\Lambda$CDM, and this requirement allows a much broader class of potentials. We return to this point in the discussion.

If the effective field theory describing cosmology is related to some underlying CFT, an interesting possibility is that this CFT and the corresponding effective theory is supersymmetric,\footnote{It has been conjectured that holographic CFTs with standard dual gravitational effective field theories must be supersymmetric \cite{Ooguri:2016pdq}, though this is not proven or widely accepted.} and that this supersymmetry is broken in the cosmology by the time-dependent scalar field expectation value.  The physics of such a model is surely very complicated. If the underlying effective theory is some gauged supergravity, the scalar expectation value would break gauge symmetry and supersymmetry, and the low energy effective field theory relevant to the cosmology would arise via RG flow in which various fields would become strongly coupled (as in QCD). We will not attempt to understand any of these things here, but ask instead a fairly naive question: can the scalar potentials giving rise to realistic time-symmetric cosmologies with asymptotically AdS Euclidean continuations arise from a superpotential in a supersymmetric effective field theory? In Section \ref{sec:susy}, we find that the answer is yes, providing an algorithm to construct a superpotential that yields the right behavior of the scalar potential for the range of scalar field values present in the cosmological solutions. 

The results of this paper are explorations to see what is possible at the level of effective field theory; we have not attempted to investigate whether the detailed potentials giving rise to realistic cosmological backgrounds can arise from microscopic constructions in string theory.  But the fact that $\Lambda < 0$ effective theories are not immediately ruled out suggests that cosmological models with asymptotically AdS continuations (and thus potential holographic descriptions) have a chance of being realistic. We discuss various directions for further investigation in Section \ref{sec:outlook}.

\section{Accelerating cosmologies from asymptotically AdS Euclidean geometries}
\label{sec:realistic}

Throughout this paper, we will consider spatially flat cosmological solutions 
\be
ds^2 = -dt^2 + a(t)^2 d \vec{x}^2 
\ee
of $\Lambda < 0$ effective field theories. For simplicity, we will consider models with a single scalar field with potential $V(\phi)$. Allowing matter and radiation, the equations of motion are 
\begin{equation}\label{eq:cosmoeom}
        \ddot{\phi} + 3 H \dot{\phi} + \frac{dV}{d\phi} = 0 \: , \qquad H^{2} = \frac{8 \pi G}{3} \left[ \frac{1}{2} \dot{\phi}^{2} + V(\phi) + \frac{\rho_{R}}{a^{4}} + \frac{\rho_{M}}{a^{3}} \right] \: ,
\end{equation}
where $H = \dot{a}/a$.

The solutions we are interested in are time-symmetric and arise via analytic continuation from real Euclidean solutions
\be
ds^2 = d \tau^2 + a_E(\tau)^2 d \vec{x}^2 
\ee
For the Euclidean solution, the equations of motion are
\begin{equation} \label{eq:wormeom}
        \phi_{E}'' + 3 H_{E} \phi_{E}' - \frac{dV}{d\phi_{E}} = 0 \: , \qquad H_{E}^{2} = \frac{8 \pi G}{3} \left[ \frac{1}{2} (\phi_{E}')^{2} - V(\phi_{E}) - \frac{\rho_{R}}{a^{4}_{E}} - \frac{\rho_{M}}{a^{3}_{E}} \right] \: ,
\end{equation}
where $a_E(\tau) = a(i \tau)$, $\phi_E(\tau) = \phi(i \tau)$, and $H_E(\tau)=H(i\tau)$.  

We would like to find potentials $V(\phi)$ and solutions $a(t), \phi(t)$ with the following properties:
\begin{itemize}
\item 
The scale factor evolution matches observational constraints, including a recent phase of accelerated expansion
\item
The solution is time-reversal symmetric about a recollapse point. 
\item
The analytic continuation of the solution (taking $t \to i \tau$ where $t=0$ is the recollapse point) is asymptotically AdS for $\tau = \pm \infty$.
\item The scalar field satisfies the Breitenlohner-Freedman (BF) bound $V'' > -9/(4L_{AdS}^2)$ at the two asymptotic AdS boundaries in the Euclidean solution.
\end{itemize}

\subsection{Constraints on the potential}

There are a variety of constraints on the scalar potential $V$ coming from these requirements.

First consider the solution in Lorentzian time (we will refer to this as the ``cosmology picture''). In order to have a time-symmetric solution, $a(t)$ and $\phi(t)$ must be even functions (taking $t=0$ to be the recollapse time, so that our present era corresponds to negative times) with $\dot{a}(0) = 0$ and $\dot{\phi}(0) = 0$. The Friedmann equation shows that the potential must be negative at this time-symmetric point, since $H=0$ here and $V$ must cancel the remaining (positive) sources of energy density.

The scalar evolution equation is that of a particle in a potential with damping $3H$ which is positive during the expanding phase. In order for the scalar field to have zero time derivative at the time-symmetric point, this particle must stop at $t=0$, and this is only possible if it is ascending the potential just prior to $t=0$. Thus, going forward from the present time, the scalar field must descend from positive to negative values of the potential and then rise and stop at 
some $\phi_0$, passing through some minimum along the way.

In the Euclidean continuation (we will refer to this as the ``wormhole picture'' since the solutions have two asymptotically AdS boundaries connected by an interior), the evolution equations are given in   (\ref{eq:wormeom}). Since $\ddot{a}(t=0) < 0$, we have $a_E''(\tau = 0) > 0$. For a solution that is asymptotically AdS as $\tau \to \pm \infty$, $a$ must behave as $e^{H_{AdS} |\tau|}$ for large $|\tau|$, so the simplest possibility is that $a$ is monotonically increasing from $\tau = 0$ towards each boundary. In this case, the evolution of $\phi$ towards the boundary is the same as the damped motion of a particle in potential $-V$. The scalar $\phi$ should approach a constant as $\tau \to \pm \infty$, so the inverted potential $-V$ should have a minimum here. Without loss of generality, we will take such a minimum to be located at $\phi=0$. 

Thus, the simplest possibility for the shape of the overall potential is that shown in Figure~\ref{fig:scalar} with the evolution in the cosmology picture and the wormhole picture covering different regions of the potential as shown. We have argued in the introduction that this shape is natural for effective field theories associated with holographic CFTs.

A final constraint comes from requiring that the AdS solution which describes the two asymptotic regions is stable to scalar perturbations. This requires that the scalar field mass (the second derivative of the potential) satisfies the BF bound
\be
\label{BF}
\left.\frac{d^2V}{d\phi^2}\right|_{\phi=0}\geq\frac{9}{4}V(0), \; \ee
where we chose units such that $8\pi G/3=1$, in which $V(0)=-1/L_{AdS}^2$. The condition (\ref{BF}) also ensures that the scaling dimensions of the corresponding operator in the dual microscopic theory is real as required for a unitary quantum field theory.

This bound turns out to have a simple implication for the scalar evolution towards the extremum at $\phi=0$. First notice that, in order to ensure the existence of the AdS asymptotic regions in the wormhole picture, the scalar potential energy term must be the dominant contribution on the right-hand side of the Euclidean Friedmann equation (\ref{eq:wormeom}) as $\tau\to\pm\infty$. This implies $H(\tau)\to H_{AdS}\equiv \sqrt{-V(0)}$ as $\tau\to\pm\infty$. By inspecting the scalar field evolution equation in (\ref{eq:wormeom}) it is then immediate to conclude that the BF bound (\ref{BF}) is equivalent to the condition for the scalar field motion to be overdamped in the vicinity of the extremum of the potential at $\phi=0$. Therefore, moving from the center of the wormhole towards the asymptotic boundaries, the scalar field will simply evolve from $\phi=\phi_0$ to $\phi=0$, settling at the minimum of the inverted potential $-V$ without any oscillation occurring in the asymptotic region.

To summarize, starting from the present value of the scalar potential, we require that it decreases to negative values and then increases to an extremum with a negative value, where the second derivative of the potential at the extremum satisfies (\ref{BF}). 

\subsection{Existence and genericity of an accelerating phase}

Starting with a potential satisfying these constraints, we can look for solutions with the desired properties.

As initial conditions at the symmetric point $t=\tau=0$, we can take\footnote{In this convention, $a$ will be less than one at the present time for examples applicable to realistic cosmology, but we can rescale $a(t) \to a(t)/a(t_0)$ to give the scale factor corresponding to the conventions of Sections \ref{sec:LambdaCDM} and \ref{sec:wcdm}.}
\be
\phi(0) = \phi_0, \qquad a(0) = 1
\ee
with vanishing derivatives $\dot a(0) = a'(0) = \dot\phi(0) = \phi'(0) = 0$.  The matter and radiation densities are constrained via the Friedmann equation at  $t=0$ as
\be
\rho_R + \rho_M  = -V(\phi_0) \; .
\ee
In the approximation where radiation can be neglected (valid for most of the cosmological history and everywhere in the wormhole solution), the matter density is fixed by $V(\phi_0)$, so the solution is fully specified by $\phi_0$, the scalar field value at $t=0$.

We would first like to check that it's possible to have an accelerating phase of cosmology before the recollapse without significant fine-tuning of the potential. Evolving backward in time from the recollapse point in the cosmology picture, the scalar field evolution corresponds to antidamped motion in the potential $V$; it is natural that the scalar reaches positive values of the potential, however this does not immediately imply an accelerating phase. Such a phase further requires the positive scalar potential to dominate over all other sources of energy density. This doesn't always happen: it may be that the growth of matter density, radiation density, or scalar kinetic energy outpaces the increase in scalar potential as $a \to 0$ or that the scale factor decreases to zero (i.e. we reach the big bang) before the potential energy dominates. Thus, it is not true that we get an accelerating phase for all potentials of the type shown in Figure \ref{fig:scalar} and all initial conditions. 

On the other hand, it is not difficult to find examples of potentials and initial conditions that do give rise to acceleration. Here we study a simple example of a potential that satisfies all of our criteria and exhibits acceleration over a significant fraction of the combined parameter space of potential parameters and the initial condition $\phi_0$. For just this subsection, we restrict to a simplified model with only radiation and scalar field contributions to the stress tensor; we set the matter contribution to zero for the purposes of this demonstration.

We consider a potential that is even about the AdS extremum $\phi = 0$ with mass $m^2 = -9/4$ at the BF bound,
\be\label{eq:exppot}
V(\phi) = -1 - \frac{9}{8} \phi^2 + V_{\text{int}}(\phi) \; .
\ee
In order to make the potential bounded from below and take positive values for larger $\phi$, we include terms at higher order in $\phi^2$. We have found that simply adding a $\phi^4$ term does not give solutions with an accelerating phase. Instead, we consider a one-parameter family of even interaction potentials 
\be\label{eq:intpot}
V_{\text{int}}(\phi) = e^{g \phi^2} - g \phi^2 - 1
\ee
where the last two terms are subtracted off to give an interaction potential starting at order $\phi^4$. 

We consider solutions for various model parameters $g$ and initial condition $\phi_0$. These solutions are obtained by numerically integrating the coupled scalar and Friedmann equations starting from the turning point where $\phi = \phi_0$ and $H=\dot{\phi}=0$. We set $\rho_M=0$ and determine the value of $\rho_R$ from $V(\phi_0)$ using~\eqref{eq:cosmoeom}. The system is numerically integrated back in time towards the big bang, then we numerically check if a given solution has a numerically detectable period of accelerated expansion.

\begin{figure}
    \centering
    \includegraphics[width=0.9\textwidth]{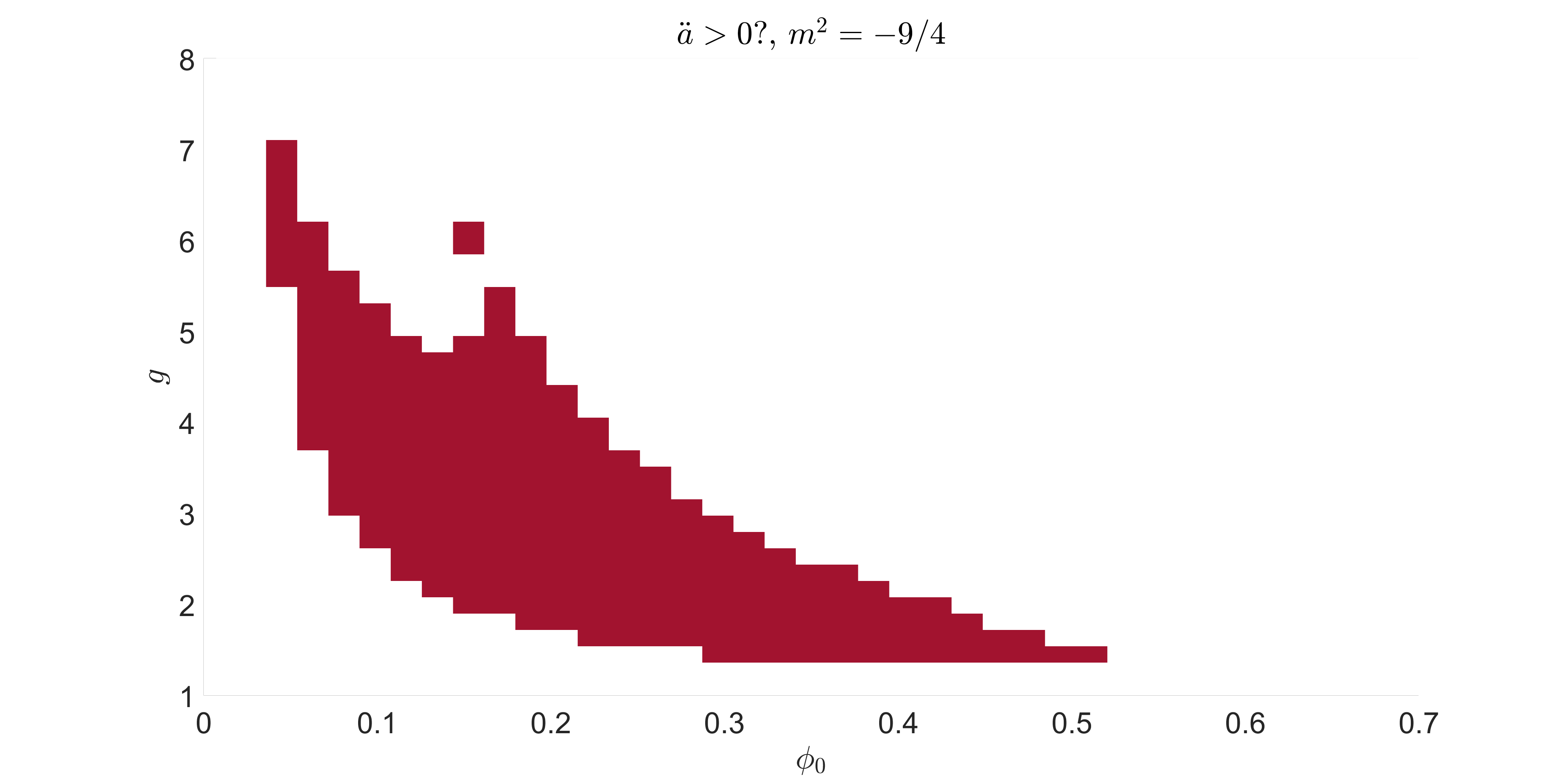}
    \caption{Space of model parameter $g$ and initial condition $\phi_0$ for the exponential potential of (\ref{eq:exppot}) and (\ref{eq:intpot}). Solutions of the equations of motion (\ref{eq:cosmoeom}) and (\ref{eq:wormeom}) with parameter $g$ and initial condition $\phi_0$ contained in the red region exhibit an accelerating phase.}
    \label{fig:paramspace}
\end{figure}

In Figure \ref{fig:paramspace}, we show the space of parameters $(g,\phi_0)$ that leads to a period of accelerated expansion in the cosmology picture before the recollapse.
We see that a significant region of the space where the parameters are of order 1 yields an accelerating phase. Thus, cosmological acceleration can be obtained without significant fine-tuning of the model or initial conditions. 

\begin{figure}
    \centering
    \includegraphics[width=0.9\textwidth]{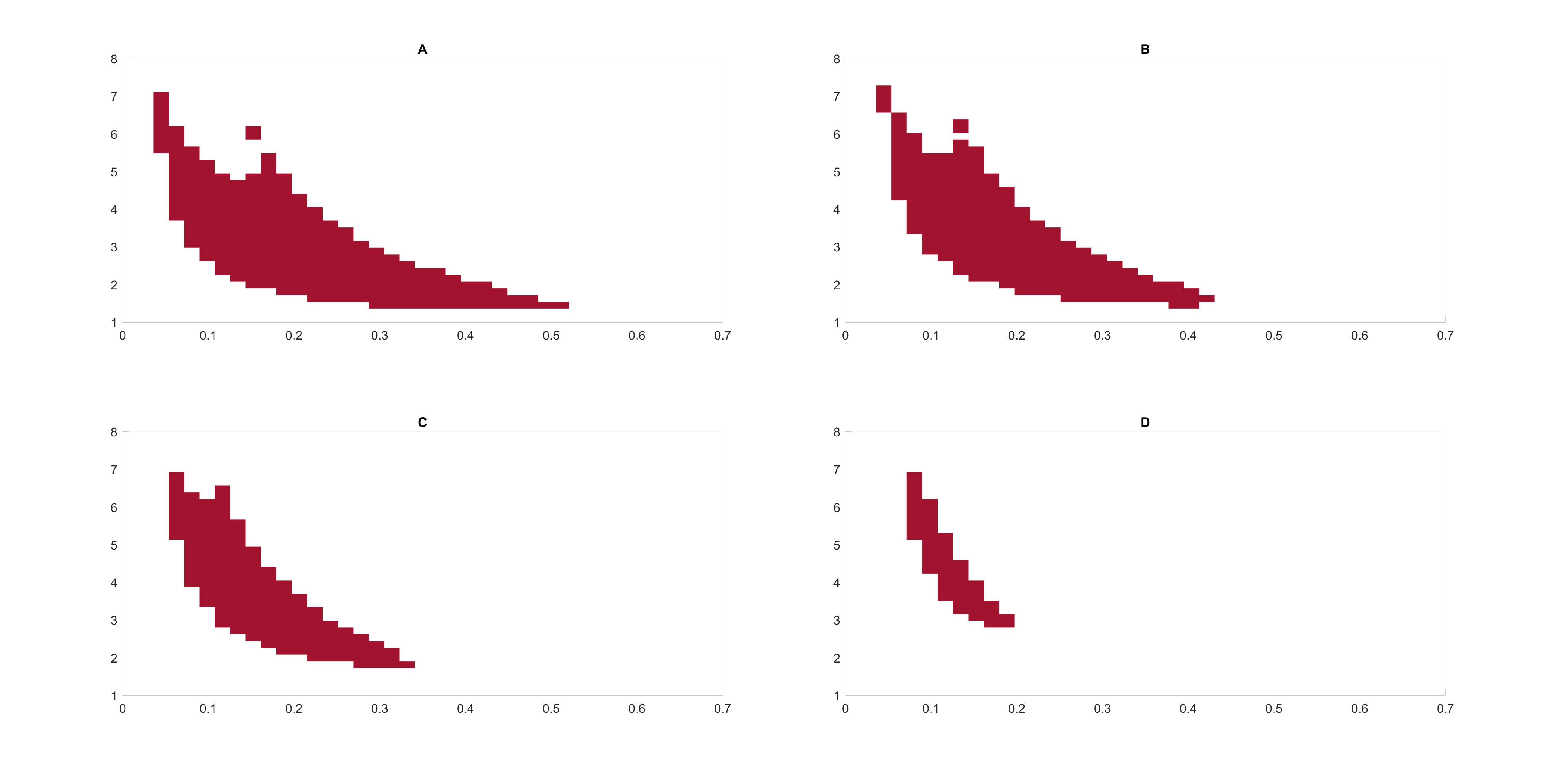}
    \caption{Panel (A) is the same as Figure~\ref{fig:paramspace} with $m^2=-9/4$. Panels (B), (C), and (D) correspond to masses $m^2=-8/4$, $m^2=-7/4$, and $m^2=-6/4$, respectively. Clearly it becomes much less generic to have a period of accelerated expansion as $m^2$ increases.}
    \label{fig:massparam}
\end{figure}

We can also consider how the mass of the field affects the allowed parameter space. In Figure~\ref{fig:massparam} we show the evolution of the parameter space exhibiting accelerated expansion as $m^2$ becomes less negative. We see that accelerated expansion becomes rapidly less generic as $m^2$ increases. Conversely, we have also observed that if $m^2$ is allowed to go below the BF bound, then a period of acceleration becomes more generic. Qualitatively, this is because the slope of the potential near the turning point (which in the present model is controlled mostly by the value of the mass) must be steep enough for the anti-damped scalar field to quickly gain kinetic energy as we evolve back towards the big bang (or forward towards the big crunch). This is necessary for the scalar field to reach the positive region of the potential before the scale factor vanishes. 

There is no particular physical significance to the form of the potential we have chosen; below, we will see that the exponentially increasing behavior of the potential for large $\phi$ is not necessary to obtain acceleration. 


\subsection{An example matching flat $\Lambda$CDM} \label{sec:LambdaCDM}


As a first example of a phenomenologically realistic cosmology based on the framework introduced in \cite{Antonini:2022blk,Antonini:2022xzo}, we study potentials able to reproduce to good accuracy the $\Lambda$CDM model. These are potentials satisfying all the properties outlined above and having a nearly flat positive region. 

For our numerical analysis it is useful to work in units such that $8\pi G/3=1$ and to introduce rescaled variables
\be
\tilde{t} = H_0 t, \qquad \Omega_i=\frac{\rho^{now}_i}{H_0^2}, \qquad \tilde{H} = \frac{H}{H_{0}}, \qquad  \tilde{V}(\phi) =\frac{V(\phi)}{H_{0}^{2}} \; ,
\label{eq:hubblerescaling}
\ee
where $\rho_i^{now}$ are the energy densities at the present time. The equations of motion then read
\begin{equation}\label{eq:cosmo_evo}
        \ddot{\phi} + 3 \tilde{H} \dot{\phi} + \frac{d\tilde{V}}{d\phi} = 0 \: , \qquad \tilde{H}^{2} = \frac{1}{2} \dot{\phi}^{2} + \tilde{V}(\phi) + \frac{\Omega_{R}}{a^{4}} + \frac{\Omega_{M}}{a^{3}} \: 
\end{equation}
in the cosmology picture, and 
\begin{equation} \label{eq:eucl_evo}
        \phi_{E}'' + 3 \tilde{H}_{E} \phi_{E}' - \frac{d\tilde{V}}{d\phi_{E}} = 0 \: , \qquad \tilde{H}_{E}^{2} =  \frac{1}{2} (\phi_{E}')^{2} - \tilde{V}(\phi_{E}) - \frac{\Omega_{R}}{a^{4}_{E}} - \frac{\Omega_{M}}{a^{3}_{E}} \: 
\end{equation}
in the wormhole picture, where overdots and primes now represent derivatives with respect to rescaled time variables. With this choice, both the Lorentzian time $\tilde{t}=H_0t$ and the Euclidean time $\tilde{\tau}=H_0\tau$ are measured in units of the Hubble time $1/H_0$. We also adopt the convention $a(\tilde{t}_0)=1$ where $\tilde{t}_0<0$ is the present time. This implies that at the time-symmetric point the initial condition for the scale factor is $a(0)=a_0>1$. The $\tilde{t}/\tilde{\tau}$ derivatives of $a$ and $\phi$ vanish at the symmetric point $\tilde{t}=\tilde{\tau}=0$. Given a value for the density parameters $\Omega_R$ and $\Omega_M$ and a choice of $a_0$, the initial condition for the scalar field is determined by the Friedmann equation at $\tilde{t}=0$:
\be
\frac{\Omega_R}{a_0^4} + \frac{\Omega_M}{a_0^3}  = -\tilde{V}(\phi_0) \; .
\label{eq:t0constraint}
\ee

\begin{figure}
    \centering
    \includegraphics[width=0.7\textwidth]{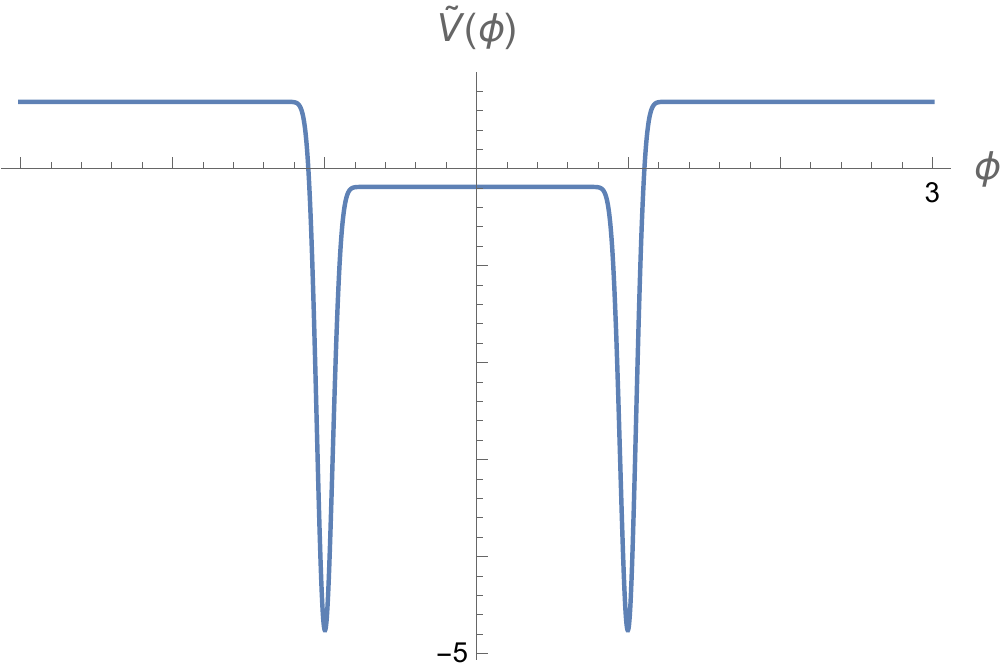}
    \caption{An example of the potential (\ref{eq:flatpot}). $A$ fixes the value of $\tilde{V}>0$ at the plateau, $B$ the value of $\tilde{V}(0)$, i.e. the negative cosmological constant in the asymptotically AdS regions of the wormhole solution. $C$, $X$, $\Delta$ determine the depth, location, and width of the deep valleys present in the potential.}
    \label{fig:flatpotential}
\end{figure}

Let us now consider a five-parameter family of potentials given by
\begin{equation} \label{eq:flatpot}
    \tilde{V}(\phi) = \tilde{V}_{0}(\phi) + \tilde{V}_{0}(-\phi) - B
\end{equation}
where
\begin{equation}\label{eq:flatpot2}
     \tilde{V}_{0}(\phi)  = \frac{A-B}{2} \textnormal{erf} \left( \frac{\phi - X}{\Delta} \right) + C \exp \left[ - \left( \frac{\phi - X}{\Delta} \right)^{2} \right] + \frac{A+B}{2}.
\end{equation}
and erf is the error function.
An example of the potential (\ref{eq:flatpot}) is plotted in Figure \ref{fig:flatpotential}. The parameter $A>0$ fixes the height of the plateau for large values of $\phi$ and determines the value of the positive vacuum energy driving the accelerated expansion in the cosmology picture, while $B<0$ gives the value of the negative cosmological constant in the asymptotic AdS regions in the wormhole picture. The parameters $C<0$, $X>0$ and $\Delta>0$ determine the depth, location, and width of the deep valleys present in the potential, respectively. In order to reproduce the $\Lambda$CDM model, the value of $A$ must match the value of the positive vacuum energy observed in our universe.

\begin{figure}
    \centering
    \includegraphics[width=0.6\textwidth]{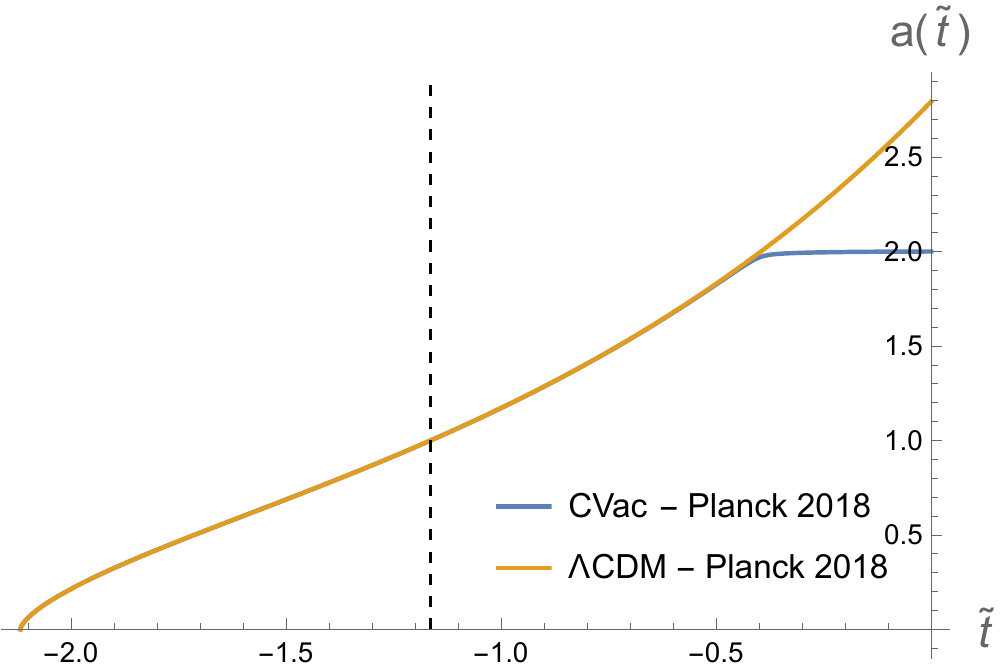}
    \caption{Scale factor evolution for the cosmological solution obtained using the \textit{Planck} cosmological parameters (\ref{eq:planckdata}) (denoted by ``CVac''). The potential parameters are given in equation (\ref{eq:numpotparam}). The scale factor for the corresponding $\Lambda$CDM solution is also depicted. The two scale factors are indistinguishable until the present day $\tilde{t}=\tilde{t}_{0}=-1.16342$ (where $a(\tilde{t}_0)=1$) --- indicated by the black dashed line --- and beyond. Deviations in our solution from the $\Lambda$CDM behavior become evident at late times as the universe approaches the turning point at $\tilde{t}=0$. The contraction phase in our solution, not depicted here, can be obtained by time-reversal.}
    \label{fig:scalecosmo}
\end{figure}

\begin{figure}
    \centering
    \includegraphics[width=0.8\textwidth]{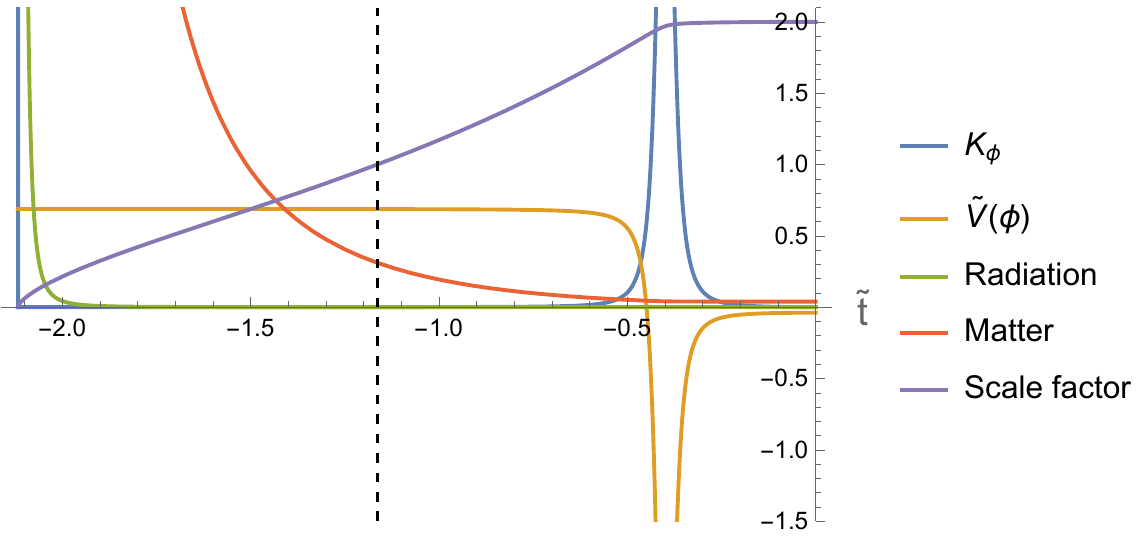}
    \caption{Scalar field kinetic and potential energies, radiation contribution $\Omega_R/a^4$, matter contribution $\Omega_M/a^3$, and scale factor $a(\tilde{t})$ as a function of time $\tilde{t}$ for the expansion phase of our cosmological solution obtained using the \textit{Planck} cosmological parameters (\ref{eq:planckdata}). The potential parameters are given in equation (\ref{eq:numpotparam}). The contraction phase can be obtained by time-reversal. The black dashed line indicates the present day $\tilde{t}=\tilde{t}_0=-1.16342$ for which $a(\tilde{t}_0)=1$. $K_{\phi}$ is negligible for most of the evolution until the present day. The universe undergoes a radiation-dominated and a matter-dominated era before the current potential energy-dominated era. In the future, the potential energy will decrease and the kinetic energy increase and become dominant as the scalar field rolls down the potential. Finally, the kinetic energy vanishes as the universe reaches its turning point at $\tilde{t}=0$, where initial conditions for our numerical solutions are imposed.}
    \label{fig:cosmoquantities}
\end{figure}

\begin{figure}
    \centering
    \includegraphics[width=0.48\textwidth]{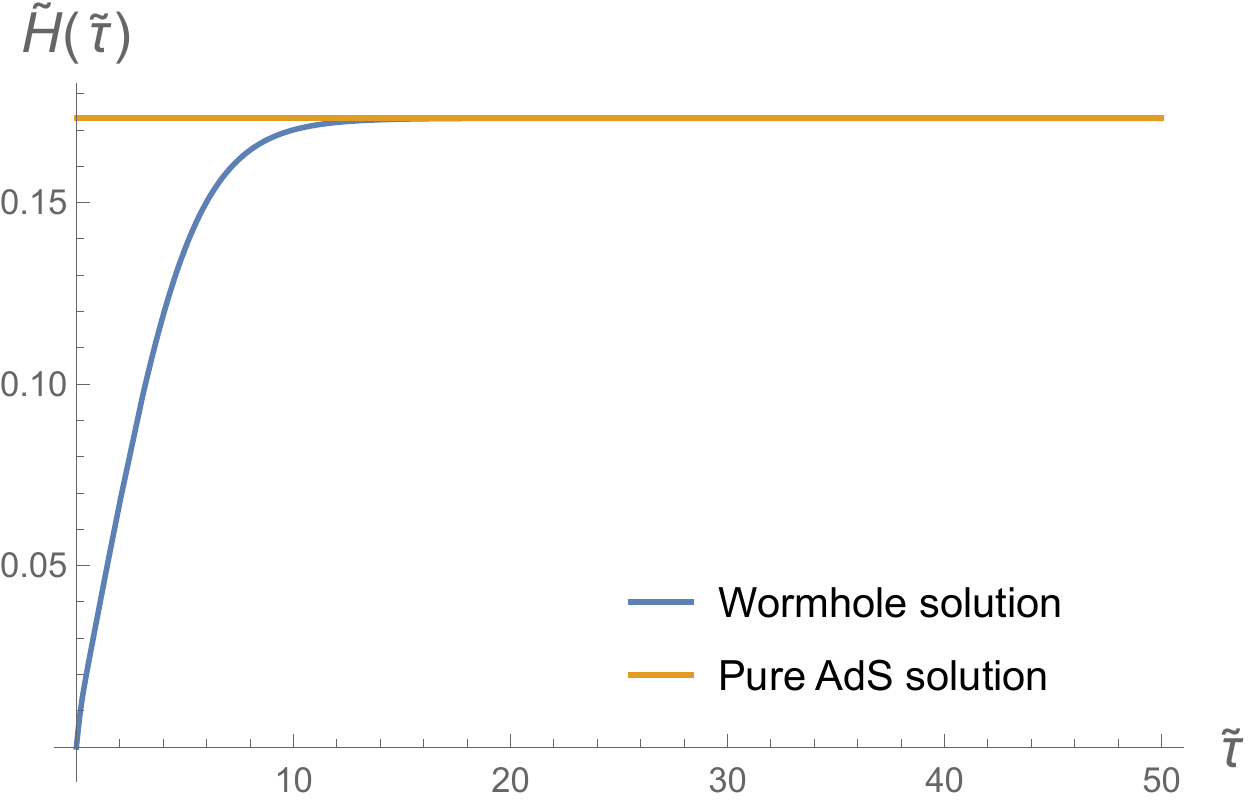}
    \includegraphics[width=0.48\textwidth]{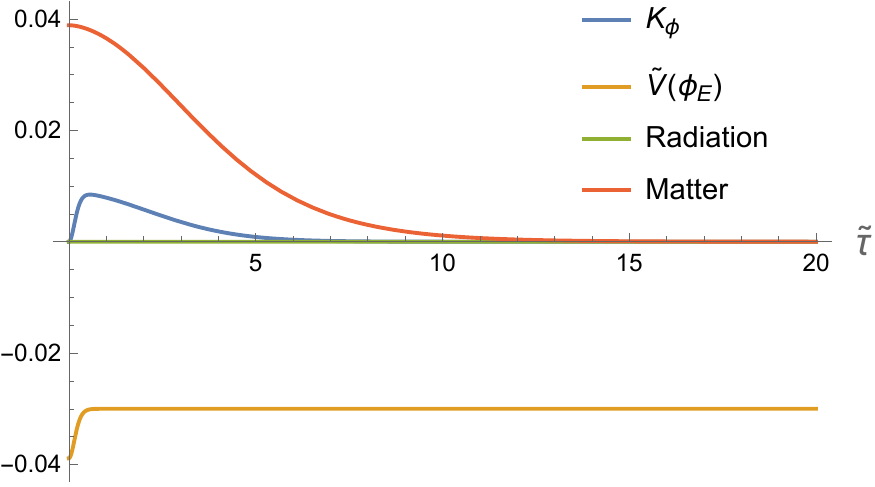}
    \caption{Wormhole solution from the center of the wormhole towards the asymptotic boundary at $\tilde{\tau}=\infty$, obtained using the \textit{Planck} cosmological parameters (\ref{eq:planckdata}). The potential parameters are given in equation (\ref{eq:numpotparam}). The evolution from the center of the wormhole towards the other asymptotic boundary at $\tilde{\tau}=-\infty$ is obtained by reflection symmetry around $\tilde{\tau}=0$. (Left) Hubble parameter as a function of the coordinate $\tilde{\tau}$. The value of the Hubble parameter for the corresponding pure AdS solution $\tilde{H}_{\infty}=\sqrt{B}$ is also depicted. In the asymptotic region $\tilde{\tau}\to\infty$ the wormhole solution approaches pure AdS. 
    (Right) Scalar field kinetic and potential energies, radiation contribution $\Omega_R/a_E^4$, and matter contribution $\Omega_M/a_E^3$ as a function of the coordinate $\tilde{\tau}$ along the wormhole direction. As we move towards the AdS asymptotic boundary (i.e. as we increase $\tilde{\tau}$), the overdamped scalar field approaches its value $\phi_E=0$ at the AdS boundary and its kinetic energy vanishes. The radiation and matter contributions are suppressed as the scale factor increases, and in the asymptotic region the negative scalar potential energy $\tilde{V}(0)$ is dominant: the solution approaches an AdS vacuum solution.}
    \label{fig:wormscale}
\end{figure}

We can then study solutions of the cosmology and wormhole equations of motion (\ref{eq:cosmo_evo}) and (\ref{eq:eucl_evo}) with potential given by equations (\ref{eq:flatpot}) and (\ref{eq:flatpot2}). We choose the values for the cosmological parameters obtained by the \textit{Planck} 2018 collaboration using TT, TE, EE + lowE + lensing + BAO data \cite{Planck:2018vyg}:
\begin{equation}
    \Omega_R=9.18\times 10^{-5}, \hspace{1cm} \Omega_M=0.311, \hspace{1cm} \Omega_{\Lambda}=0.6889
    \label{eq:planckdata}
\end{equation}
where $\Omega_\Lambda$ is the density parameter associated with the vacuum energy today. We further choose the initial condition $a_0=2$ for the scale factor at the time-symmetric point, while the scalar field initial condition $\phi_0$ is determined by equation (\ref{eq:t0constraint}) once the five parameters of the potential are fixed.\footnote{We remind that the initial conditions for the cosmology and wormhole pictures are the same. Note also that in general there are multiple solutions of equation (\ref{eq:t0constraint}). As we have explained above, we are interested in a solution such that $\phi_0$ is in between a minimum and a maximum of the potential.} In our numerical solutions depicted in Figures \ref{fig:scalecosmo}, \ref{fig:cosmoquantities}, and \ref{fig:wormscale}, they are taken to be
\begin{equation}
    A=0.6889, \hspace{0.7cm} B=-0.03, \hspace{0.7cm} C=-5, \hspace{0.7cm} X=1, \hspace{0.7cm} \Delta=0.0726775778709
    \label{eq:numpotparam}
\end{equation}
yielding $\phi_0=0.817319$.

\paragraph{Cosmological solution}

By an appropriate fine-tuning of the potential's parameters (see Appendix \ref{app:potentials}) leading to the values (\ref{eq:numpotparam}), the cosmological evolution reproduces to very good accuracy the one predicted by the corresponding $\Lambda$CDM solution between the early universe and the present day $\tilde{t}=\tilde{t}_0$, see Figure \ref{fig:scalecosmo}. For $\tilde{t}>\tilde{t}_0$, the potential energy eventually decreases and becomes negative as the scalar field rolls down the potential, while the kinetic energy $K_{\phi}$ increases. Finally, the scalar kinetic energy decreases and vanishes as we reach the time-symmetric point $\tilde{t}=0$ (where we impose our initial conditions for both the cosmological and the wormhole solutions). See Figure \ref{fig:cosmoquantities}.

\begin{figure}
    \centering
    \includegraphics[width=\textwidth]{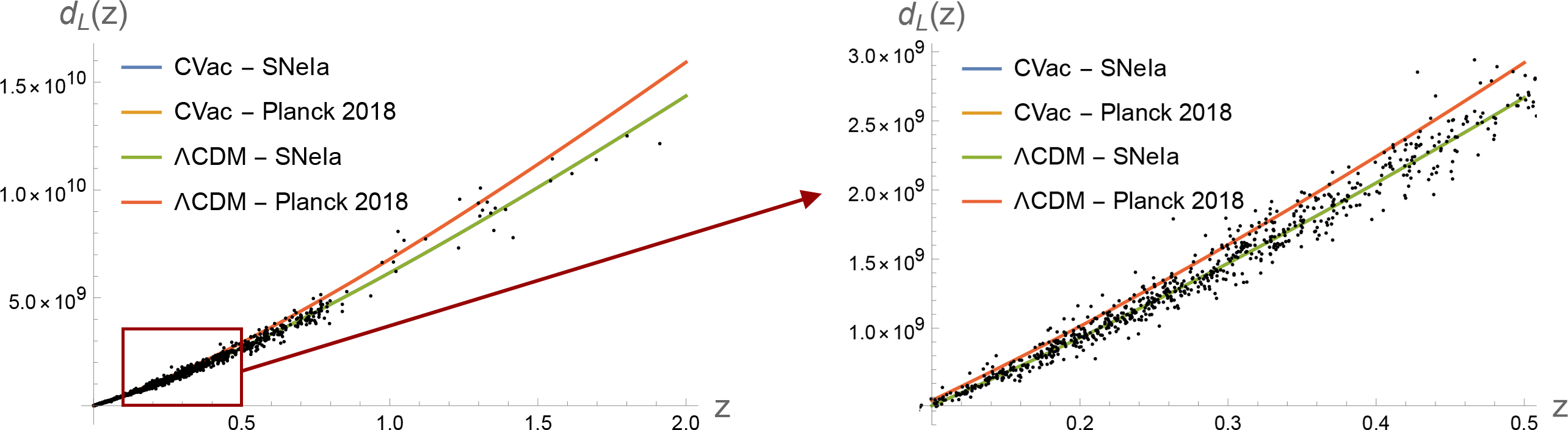}
    \caption{Luminosity distance $d_L(z)$ computed for two solutions involving rolling scalars (denoted by ``CVac'') and their corresponding $\Lambda$CDM solutions. For the \textit{Planck} 2018 solution, the cosmological parameters are given in equation (\ref{eq:planckdata}), the potential parameters in equation (\ref{eq:numpotparam}), and we used $H_0^{Planck}=67.66\textrm{ km s}^{-1}\textrm{ Mpc}^{-1}$ \cite{Planck:2018vyg}. For the SNeIa solution, the cosmological parameters are given by\protect\footnotemark~$\Omega_R=9.96\times 10^{-5}$, $\Omega_M=0.338$, $\Omega_\Lambda=0.662$, $H_0^{SN}=73.4\textrm{ km s}^{-1}\textrm{ Mpc}^{-1}$ \cite{Brout:2022vxf}, and the potential parameters are $A=0.662$, $B=-0.03$, $C=-5$, $X=1$, $\Delta=0.0716914850735$; this yields $\phi_0=0.824448$. The Pantheon+SH0ES experimental data are also depicted \cite{Pantheondata}. Our cosmological solutions and their corresponding $\Lambda$CDM solutions are indistinguishable, meaning that our model matches supernovae data as well as the $\Lambda$CDM model. Notice that the solutions generated using \textit{Planck} 2018 cosmological parameters are in tension with data, while the ones generated using cosmological parameters derived from supernovae observations agree with data: this is a manifestation of the Hubble tension.}
    \label{fig:muflatpot}
\end{figure}

As a consistency check, we verified that the luminosity distance 
\begin{equation}
    d_L(z)=\frac{1}{H_0a(t)} \int_{\tilde{t}}^{\tilde{t}_0} \frac{dt'}{a(t')}  \: , \qquad 1 + z = \frac{1}{a(t)} \: 
\end{equation} 
computed from our solution\footnote{For the solution obtained using the \textit{Planck} cosmological parameters (\ref{eq:planckdata}) we can use the value of $H_0$ measured by \textit{Planck}, i.e. $H_0^{planck}=67.66\textrm{ km s}^{-1}\textrm{ Mpc}^{-1}$.} agrees with the Pantheon+SH0ES type Ia supernova (SN Ia) data \cite{Pantheondata}. In particular, we found that the luminosity distance in our solution is indistinguishable from the one generated by the corresponding $\Lambda$CDM solution in the range of redshifts $z$ covered by the SNIa data, see Figure \ref{fig:muflatpot}. Therefore, our model fits the data as well as the $\Lambda$CDM model. Due to the well-known Hubble tension issue \cite{DiValentino:2021izs}, this also implies that the solution obtained using the \textit{Planck} 2018 parameters (\ref{eq:planckdata}) is in tension with supernovae data. We also studied a solution generated using cosmological parameters obtained from supernovae observations \cite{Brout:2022vxf}, which clearly matches data much better. Whether specific realizations of our models could help solve the Hubble tension is an interesting open question that we plan to investigate in future work.

\paragraph{Wormhole solution}

First, we remark that the potential under consideration with the parameters (\ref{eq:numpotparam}) used in our numerical analysis satisfies the BF bound (\ref{BF}). As a result of the scalar field's overdamped motion described above, the scale factor in the wormhole solution $a_E(\tilde{\tau})$ increases away from the wormhole center, the matter and radiation terms are suppressed as $\tilde{\tau}$ increases or decreases from $\tilde{\tau}=0$, and the scalar kinetic energy $K_{\phi}$ becomes negligible as $\phi_E\to 0$ (see Figure \ref{fig:wormscale}, right panel). Therefore, for $\tilde{\tau}\to \pm \infty$ the solution approaches pure AdS with vacuum energy density parameter given by $\Omega_{\Lambda}=3\tilde{V}(0)=3 B$. As a consequence, the scale factor takes the asymptotic form $a(\tilde{\tau})\to \exp(\sqrt{-B}\tilde{\tau})$ as $\tau\to\pm\infty$ and the Hubble parameter approaches a constant value $\tilde{H}(\tilde{\tau})\to \tilde{H}_{\infty}=\sqrt{-B}$ (see Figure \ref{fig:wormscale}, left panel). These results confirm the existence, within the class of models introduced in \cite{Antonini:2022blk,Antonini:2022xzo}, of well-defined wormhole solutions associated with cosmological solutions able to reproduce to arbitrary accuracy the predictions of the $\Lambda$CDM model.

\footnotetext{In the late stages of this manuscript's preparation a revised version of \cite{Brout:2022vxf} was announced, with a slight correction of the best fits for the cosmological parameters, yielding $\Omega_M=0.334\pm 0.018$ and $\Omega_{\Lambda}= 0.666 \pm 0.018$. This difference is irrelevant for the purposes of the present paper.}


We would like to point out that, from a top-down point of view, there is reason to be skeptical about the existence of the type of potentials studied in this subsection on the grounds of Swampland conjectures. In particular, the de Sitter conjecture \cite{Obied:2018sgi, Garg:2018reu, Ooguri:2018wrx} would appear to rule out potentials with very flat regions such as the ones introduced in equations (\ref{eq:flatpot}) and (\ref{eq:flatpot2}). Nonetheless, this class of nearly-flat potentials is a useful starting point for our discussion. Indeed, the present analysis suggests that sufficient fine-tuning allows us to reproduce arbitrarily well the predictions of the $\Lambda$CDM model between the early universe and the present time while preserving the existence of a wormhole solution with the scalar field satisfying the BF bound at the two asymptotic AdS boundaries. A similar conclusion remains valid for a more general class of potentials, as we will see in the next subsection.

\subsection{An example matching flat \texorpdfstring{$w$}{}CDM} \label{sec:wcdm}



Though we have considered a flat potential to make direct contact with the $\Lambda$CDM model, such tuning of the scalar potential is not at all necessary, and we can consider a broader class of models consistent with direct observations of $\Omega_{M}$ and the scale factor $a(t)$, which may involve scalar evolution over a range of potential values $\Delta V / V = O(1)$.

For concreteness, we consider the spatially flat $w$CDM model,\footnote{We neglect radiation in this subsection.} since direct constraints on the cosmological parameters for these models from type Ia supernova observations, using the Pantheon+SH0ES data set, are provided in \cite{Brout:2022vxf}. We recall that this model consists of modifying the dark energy contribution in the $\Lambda$CDM model, by replacing the cosmological constant with a perfect fluid governed by the equation of state $\rho = w p$, with $w$ constant. The corresponding Hubble expansion is given by
\begin{equation} \label{eq:H_w}
    H(z) = H_{0} \sqrt{\Omega_{M} (1+z)^{3} + \Omega_{\Lambda}(1+z)^{3(1+w)}} \: ,
\end{equation}
where $\Omega_{\Lambda}$ is the density parameter for the dark energy, so that $\Omega_{\Lambda} = 1 - \Omega_{M}$ with the assumption of spatial flatness. The constraints on the parameters of this model from \cite{Brout:2022vxf} are 
\begin{equation}
    H_{0} = (73.5 \pm 1.1) \: \textnormal{km} \cdot \textnormal{s}^{-1} \cdot \textnormal{Mpc}^{-1} \: , \quad \Omega_{M} = 0.309^{+0.063}_{-0.069} \: , \quad w = -0.90 \pm 0.14 \: .
\end{equation}

\begin{figure}
    \centering
    \includegraphics[width = 0.45\textwidth]{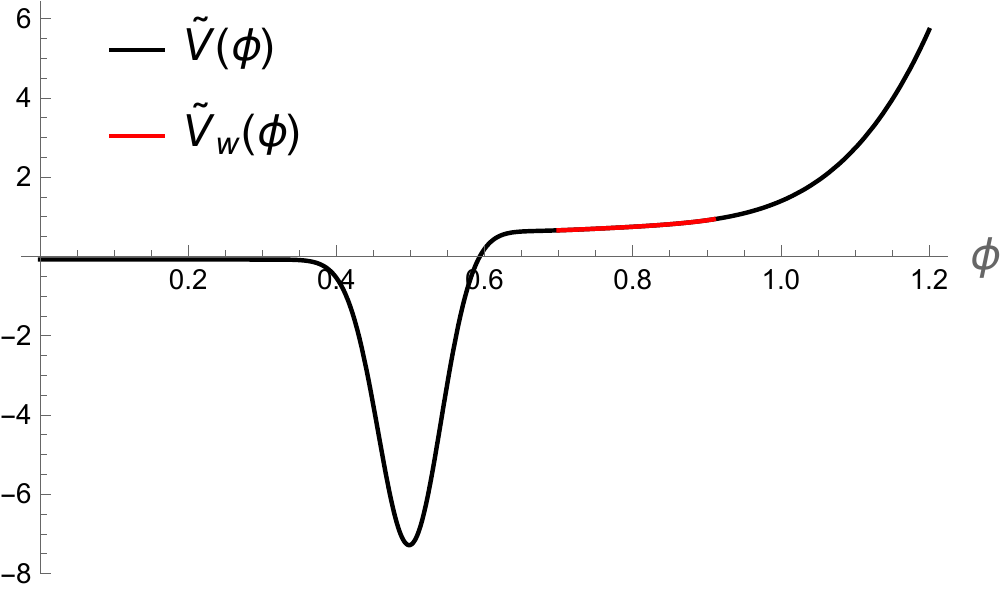}
    \includegraphics[width = 0.45\textwidth]{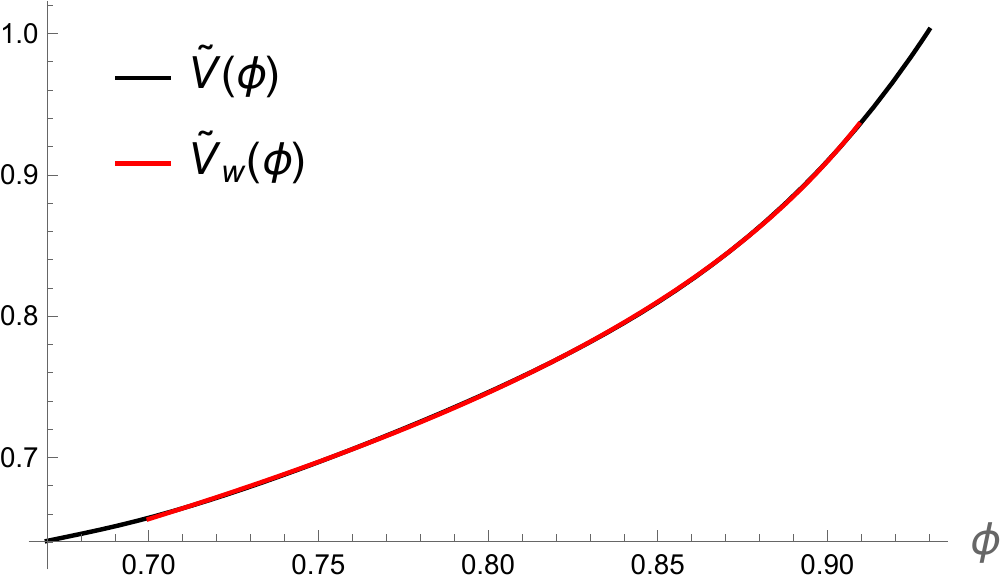}
    \caption{Rescaled potentials for the model $V(\phi)$ and reconstruction $V_{w}(\phi)$, where the region of the latter probed by the scalar field in the range $z \in (z_{\textnormal{min}}, z_{\textnormal{max}})$ is shown. The righthand plot is a close-up on this region. The precise form of the model $V(\phi)$ can be found in Appendix \ref{app:wcdm}. }
    \label{fig:2022_Nov16_wCDM_pot}
\end{figure}

We can reproduce the evolution of the $w$CDM cosmology over an arbitrarily large range of redshifts using our scalar field model. With the assumption that only a single scalar field is relevant to the cosmological evolution, we can actually deduce the potential associated with this scalar field given $a(t)$ and $\Omega_M$. The kinetic energy can be expressed as
\be
K(t) \equiv {1 \over 2} \dot{\phi}^2 = -{1 \over 8 \pi G} \dot{H} -{3 \over 16 \pi G} \Omega_M H_0^2 {1 \over a^3} \;,
\ee
and we can take the square root and integrate, to obtain
\be
\label{phisol}
\phi(t) = \phi(t_0)+\int_{t_0}^{t} d \hat{t} \sqrt{-{1 \over 4 \pi G} \dot{H}(\hat t) -{3 \over 8 \pi G} \Omega_M H_0^2 {1 \over a^3(\hat t)}}.
\ee
Similarly, the potential energy is 
\be
\label{V3}
V(t) = {1 \over 8 \pi G} \dot{H} +  {3 \over 8 \pi G} H^2 -{3 \over 16 \pi G} \Omega_M H_0^2 {1 \over a^3} \;.
\ee
We can alternatively express these in terms of the redshift $z$ and adopt the rescaled quantities introduced in the previous subsection, obtaining
\begin{equation} \label{eq:Hz_phiz_wCDM}
    \begin{split}
        \phi(z) & = \phi_{0} + \int_{0}^{z} dz' \: \frac{\sqrt{\frac{1}{4 \pi G}(1+z') \tilde{H}(z') \tilde{H}'(z') - \frac{3}{8 \pi G} (1+z')^{3} \Omega_{M}}}{(1+z') \tilde{H}(z')} \\
        \tilde{V}(z) & = - \frac{1}{8 \pi G} (1+z) \tilde{H}(z) \tilde{H}'(z) + \frac{3}{8 \pi G} \tilde{H}(z)^{2} - \frac{3}{16 \pi G} (1+z)^{3} \Omega_{M} \: .
    \end{split}
\end{equation}
We may then substitute the rescaled version of the Hubble expansion from equation (\ref{eq:H_w}) into these expressions. 

In the scalar field model, the effective equation of state parameter is given by
\begin{equation} \label{eq:w}
    w = \frac{\frac{1}{2} \dot{\phi}^{2} - V}{\frac{1}{2} \dot{\phi}^{2} + V} \: .
\end{equation}
For a general scalar theory, $w$ will be redshift dependent, but with appropriate choice of scalar potential, we may recover solutions for which $w$ is approximately constant over an arbitrarily large range of redshifts. We denote by $V_{w}(\phi)$ the scalar potential reconstructing the $w$CDM model for all $z>0$. 
For the best-fit parameters of $w$CDM listed above, we show part of the reconstructed potential $\tilde{V}_{w}(\phi)=V_{w}(\phi)/H_0$ in Figure \ref{fig:2022_Nov16_wCDM_pot}. 

Since the Pantheon+SH0ES data used to deduce the cosmological parameter constraints probe only redshifts in the range $z \in (z_{\textnormal{min}}, z_{\textnormal{max}})$ with $z_{\textnormal{min}} \approx 10^{-3}$ and $z_{\textnormal{max}} \approx 2.26$, 
we are free to consider any scalar potential $V(\phi)$ which coincides with $V_{w}(\phi)$ in the region over which the scalar solution evolves in this range of redshifts. 
In Figure \ref{fig:2022_Nov16_wCDM_pot}, we also plot a rescaled scalar potential $\tilde{V}(\phi)=V(\phi)/H_0$ agreeing with $\tilde{V}_{w}(\phi)$ in the appropriate range, but that satisfies the ``UV constraints" necessary for the existence of an analytic continuation of the cosmological solution to an asymptotically AdS wormhole. The precise form of the potential $\tilde{V}(\phi)$, as well as our method for constructing the potential and the appropriate initial conditions for the cosmological solution, are given in Appendix \ref{app:wcdm}. In the left panel of Figure \ref{fig:2022_Nov16_wCDM_sol}, we plot the equation of state evolution $w(z)$ extracted directly from the time-symmetric cosmological solutions in the model with potential $\tilde{V}(\phi)$, confirming that they reproduce that of the $w$CDM model with the desired precision, while in the right panel of Figure \ref{fig:2022_Nov16_wCDM_sol} we confirm that the analytic continuation of these solutions have AdS asymptotics. 

We emphasize that, while we have only aimed here to reproduce the $w$CDM model over the region $z \in (z_{\textnormal{min}}, z_{\textnormal{max}})$ where it has been constrained by SNe Ia data, we could recover the same model over an arbitrarily large redshift interval through suitable choice of the scalar potential.


\begin{figure}
    \centering
    \includegraphics[width = 0.45\textwidth]{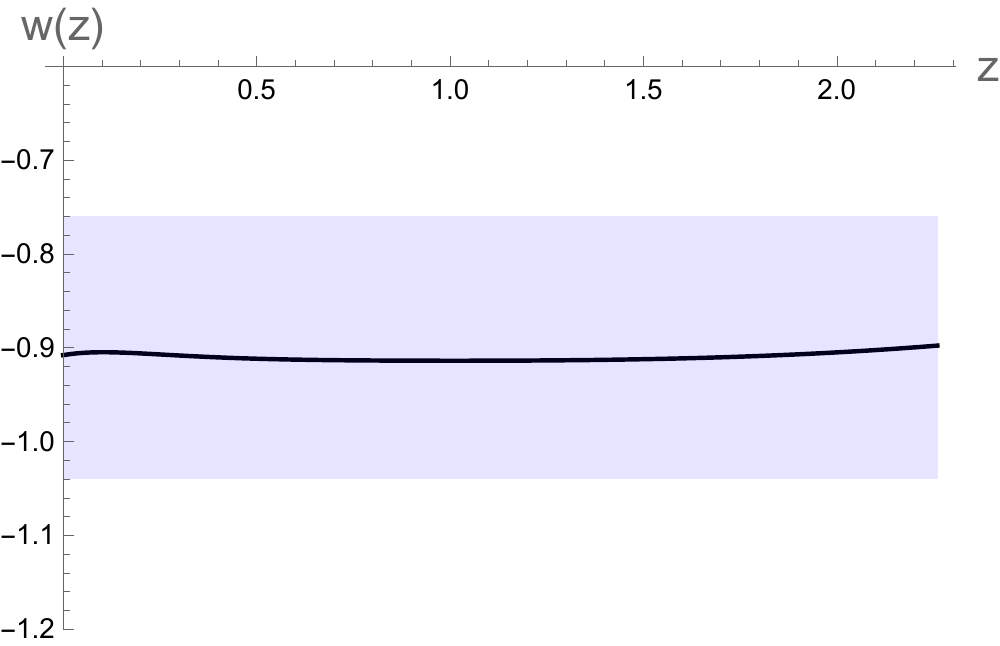}
    \includegraphics[width = 0.45\textwidth]{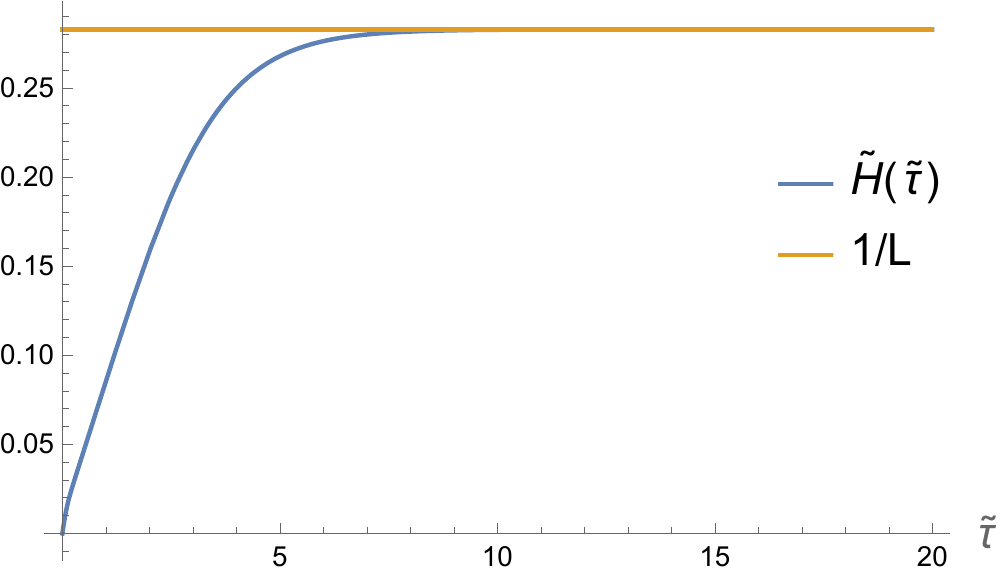}
   \caption{(Left) Equation of state parameter $w$ versus redshift $z$ for time-symmetric solutions to the model $\tilde{V}(\phi)$, with shaded region given by the Pantheon+SH0ES constraint. (Right) Euclidean evolution of $\tilde{H}(\tilde{\tau})$, with $\frac{1}{L} \equiv \sqrt{- V(0)}$ plotted for reference. In the asymptotic region $\tilde{\tau}\to\infty$ the wormhole solution approaches pure AdS, as expected. The precise form of the potential $\tilde{V}(\phi)$ can be found in Appendix \ref{app:wcdm}. }
    \label{fig:2022_Nov16_wCDM_sol}
\end{figure}

\section{Supersymmetric models}
\label{sec:susy}

In our discussion so far, we have considered an effective gravitational field theory involving scalar fields with dynamics controlled by an effective potential $V$. We have shown that such a theory could describe the background dynamics of our Universe while also admitting a Euclidean continuation suggestive of a dual CFT description.

The most well-understood examples of gravitational theories which are dual to CFTs are supersymmetric. It is thus interesting to ask whether there exist supersymmetric theories whose effective low energy descriptions contain scalar fields with potentials $V$, which give rise to effective cosmological dynamics. In these examples, supersymmetry could be broken by the time-dependent scalar field expectation value. Note that here we will not be concerned with whether the supersymmetric theories under consideration have explicitly known CFT duals. Instead we simply wish to understand whether supersymmetry is consistent with the restricted class of scalar potentials which we have considered so far.

We will consider $\mathcal N = 1$ supergravity and restrict to considering only the gravity and scalar sectors; presumably one can consistently set the gravitino and chiral fermions to zero in solutions to the full equations of motion. We follow the notation of \cite{deAlwis:2013jaa} and we set $8 \pi G_{\textnormal{N}} = \kappa^{2} = 1$ in this section.

The $n_c$ scalar fields $\phi^{i}$ of $\mathcal{N}=1$ supergravity are bottom components of chiral multiplets, such that the theory has bosonic Lagrangian
\begin{equation}
    \mathcal{L} = \frac{1}{2} R - K_{i \bar{j}} \partial_{\mu} \phi^{i} \partial^{\mu} \bar{\phi}^{\bar{j}} - V_W(\phi, \bar{\phi}) \: ,
\end{equation}
where $K_{i \bar{j}} = \partial_{i} \bar{\partial}_{\bar{j}} K$ is the K{\"a}hler metric arising from a K{\"a}hler potential $K$, and $V_W$ is the scalar potential
\begin{equation}
    V_W = e^{K} \left( K^{i \bar{j}} D_{i} W D_{\bar{j}} \bar{W} - 3 |W|^{2} \right) \: , \qquad D_{i} W = \partial_{i} W + K_{i} W \: ,
\end{equation}
defined in terms of the superpotential $W$.

For simplicity we will consider the case of a single chiral multiplet, $n_c = 1$, and set the K{\"a}hler potential to the canonical form $K(\phi, \bar{\phi}) = \bar{\phi} \phi$. Theories with more general K{\"a}hler potentials and more chiral multiplets may also be interesting to consider. In our case the scalar potential is given by
\begin{align}\label{eq:V_W_general}
    V_W(\phi) = e^{\phi\bar\phi}\left(|W'(\phi)+\bar\phi W(\phi)|^2 -3|W(\phi)|^2\right).
\end{align}
A specification of $W$ is part of the specification of the supersymmetric theory, with the only restriction on $W$ being that it is a holomorphic function of $\phi$. Thus we would like to know whether it is possible to choose $W$ such that $V_W$ is a scalar potential like the ones considered in previous sections, i.e. which gives rise to effective cosmological dynamics.

The scalar field $\phi$ arising from the supersymmetric theory is a complex field, which we can decompose into two real fields as $\phi = \phi_R + i \phi_I$. Thus $V_W$ is a (real) potential of two real fields, $\phi_R$ and $\phi_I$, whose dynamics will in general be coupled --- the trajectory of $\phi$ in the complex plane will generally involve motion in both the $\phi_R$ and $\phi_I$ directions.

However, although it is certainly possible that the low energy effective field theory describing our Universe contains multiple real scalar fields, in the previous sections we have shown that a single field with potential $V$ is sufficient for the cosmology to be in quantitative agreement with measurements of the scale factor, while still admitting an asymptotically AdS Euclidean continuation. Thus we might hope that we can find a potential $V_W$ such that the dynamics in the $\phi$ plane is one-dimensional, for example such that $\phi$ rolls down the real axis, with $\phi_I = 0$ for all times. Moreover we would like $V_W(\phi_R) = V(\phi_R)$, so that the remaining scalar degree of freedom $\phi_R$ provides us with the desired cosmological evolution. 

We will now show that, even with this restriction that $\phi_I$ effectively decouples from the dynamics, given any cosmological potential $V(\phi_R)$ it is possible to find a superpotential $W(\phi)$ such that $V_W(\phi_R)$ is, to an arbitrarily good approximation, equal to $V(\phi_R)$ within a given fixed region. Furthermore, in this construction $W$ will be holomorphic everywhere provided that $V$ is. Since the cosmological potentials which we considered in previous sections are generically holomorphic everywhere, this construction will produce holomorphic superpotentials $W$ which give rise to realistic effective cosmological dynamics with Euclidean AdS asymptotics. 

\paragraph{Constructing the superpotential}

In our construction $W(\phi_R)$ will be real, in which case Eq. \eqref{eq:V_W_general} for $\phi = \phi_R$ reads
\begin{align}
    V_W(\phi_R) 
    &= 
    e^{\phi_R^2}
    \left(\left[W'(\phi_R)+\phi_R W(\phi_R)\right]^2 -3W^2(\phi_R)\right).
\end{align}
This is a nonlinear ODE for $W$ and cannot be solved in closed form. To obtain a linear equation, write
\begin{align}\label{eq:V_W_factored}
    V_W(\phi_R)
    &=
    e^{\phi_R^2}
    \left( 
    W'+(\phi_R+\sqrt{3})W
    \right)
    \left( 
    W'+(\phi_R-\sqrt{3})W
    \right).
\end{align}
Now let $f(\phi_R)$ be some function, which we define shortly, and consider the linear ODE
\begin{align}
    W'(\phi_R)+(\phi_R-\sqrt{3})W(\phi_R) = f(\phi_R).
\end{align}
The general solution is the sum of a homogeneous plus inhomogeneous piece,
\begin{align}\label{eq:W_first_equation}
    W(\phi_R) = C G(\phi_R)+ G(\phi_R)\int_0^{\phi_R} du \frac{f(u)}{G(u)},
\end{align}
where $C$ is an arbitrary integration constant, and the homogeneous piece is given by
\begin{align}
    G(\phi_R) = \exp\left(-\frac{1}{2}\phi_R^2+\sqrt{3}\phi_R\right).
\end{align}
Substituting this solution for $W$ into \eqref{eq:V_W_factored} we obtain the scalar potential
\begin{align}\label{eq:V_W}
    V_W(\phi_R) = 
    \tilde V(\phi_R)
    \left(
    1+\frac{\tilde V(\phi_R) e^{-2\sqrt{3}\phi_R}}{12C^2}
    +\frac{1}{2\sqrt{3}C^2} \int_0^{\phi_R} du \,\tilde V(u) e^{-2\sqrt{3}u}
    \right).
\end{align}
where we have defined
\begin{align}\label{eq:f}
    \tilde V(\phi_R) \equiv 2\sqrt{3} C e^{\phi_R^2} f(\phi_R)G(\phi_R).
\end{align}

Because we are free to choose $f$, we are effectively free to choose $\tilde V$. We can thus require $V_W(\phi_R) = V(\phi_R)$ and solve \eqref{eq:V_W} for $\tilde V$. Although this cannot be done in closed form, we can obtain a series approximation for $\tilde V$ in powers of $1/C^2$. Notice that $C$ is an arbitrary constant, and so it can be chosen large enough to be a good expansion parameter; we will discuss this in more detail below. Namely, writing
\begin{align}
    \tilde V(\phi_R) = \tilde V_0(\phi_R) + \frac{1}{C^2}\tilde V_2(\phi_R)+ \frac{1}{C^4}\tilde V_4(\phi_R)+\dots,
\end{align}
it is possible to recursively solve the equation $V_W(\phi_R) = V(\phi_R)$ for the $\tilde V_n$. The first two terms in the expansion are
\begin{align}
    \tilde V_0(\phi_R) &= V(\phi_R),\\
    \tilde V_2(\phi_R) &= -\frac{V(\phi_R)}{2\sqrt{3}}
    \left(\frac{e^{-2\sqrt{3}\phi_R}V(\phi_R)}{2\sqrt{3}}+\int_0^{\phi_R} du\,V(\phi_R)e^{-2\sqrt{3}u}\right),
\end{align}
and expressions for higher order terms can be straightforwardly obtained by recursion. Substituting \eqref{eq:f} into \eqref{eq:W_first_equation} we obtain the superpotential in terms of $\tilde V$,
\begin{align}
    W(\phi_R) = C G(\phi_R)
    \left(
    1+\frac{1}{2\sqrt{3}C^2}\int_0^{\phi_R} du \, \tilde V(u) e^{-2\sqrt{3}u} 
    \right),
\end{align}
and using the series expansion for $\tilde V$ in terms of $V$, we can, perturbatively in powers of $1/C^2$, write the superpotential $W$ in terms of the cosmological potential $V$:
\begin{align}
    W(\phi_R) = C G(\phi_R)
    \left[
    1+\frac{1}{2\sqrt{3}C^2}\int_0^{\phi_R} du \, V(u) e^{-2\sqrt{3}u}
    + \mathcal{O}\left(C^{-4}\right)\right] .
\end{align}

\paragraph{Truncating the expansion}

If we use the full expression for $W(\phi_R)$ derived above, including all corrections in powers of $1/C^2$, then we would expect to find an exact equality $V_W(\phi_R) = V(\phi_R)$ for all values of $\phi_R$. Of course, it is not possible to work with the infinite series of terms. Instead we can truncate the series expansion of $W$ at $n$ terms, obtaining the truncated superpotential, which we denote $W_n$. For example, $W_2$ is obtained by keeping two terms in the expansion,
\begin{align}\label{eq:W2}
    W_2(\phi_R) = C G(\phi_R)
    \left(
    1+\frac{1}{2\sqrt{3}C^2}\int_0^{\phi_R} du \, V(u) e^{-2\sqrt{3}u}\right),
\end{align}
which results in the scalar potential
\begin{align}\label{eq:V_W2}
    V_{W_2}(\phi_R) = 
    V(\phi_R)
    \left(
    1+\frac{V(\phi_R) e^{-2\sqrt{3}\phi_R}}{12C^2}
    +\frac{1}{2\sqrt{3}C^2} \int_0^{\phi_R} du \,V(u) e^{-2\sqrt{3}u}
    \right).
\end{align}

Recall that $C$ appeared simply as an undetermined integration constant, and we are free to set it to any value whatsoever. Therefore we have the somewhat unusual freedom to set our expansion parameter $1/C$, and thus the terms truncated in going from $W$ to $W_2$, as small as we like. In terms of the potentials, given any value of $\phi_R$ we can make $V_{W_2}(\phi_R)$ arbitrarily close to the cosmological potential $V(\phi_R)$. In other words, as we take $C$ large $V_{W_2}$ converges pointwise to $V(\phi_R)$. If we restrict to values of $\phi_R$ in any compact subset of the real line, then the convergence is uniform within this subset. Thus we have successfully managed to embed any scalar potential $V$ --- including cosmological potentials --- into a supersymmetric theory, to an arbitrary degree of accuracy.

\begin{figure}
    \centering
    \includegraphics[width = 0.7\textwidth]{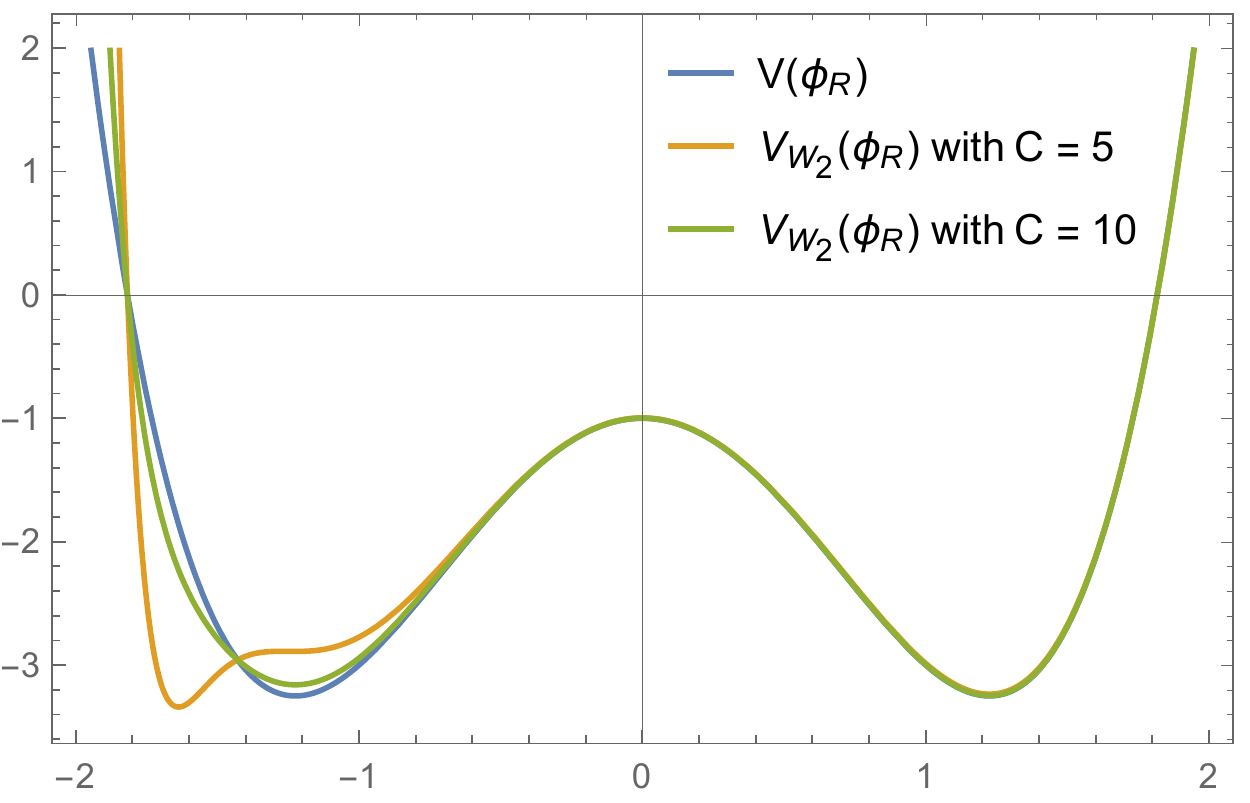}
    \caption{Comparison of $V(\phi_R) = -1 -3\phi_R^2 + \phi_R^4$ and the potential $V_{W_2}(\phi_R)$ given by equation (\ref{eq:V_W2}), obtained from the truncated superpotential $W_2$ given by equation (\ref{eq:W2}).}
    \label{fig:V_and_VW2}
\end{figure}

In Fig. \ref{fig:V_and_VW2} we illustrate with a simple example the deviation of $V_{W_2}$ from $V$ for various values of $C$. Notice that $V_{W_2}(\phi_R)$ is in excellent agreement with $V(\phi_R)$ for $\phi_R>0$, but for a fixed value of $C$ the agreement breaks down for negative enough values of $\phi_R$. This is because the exponential factor in the integrand in \eqref{eq:W2} suppresses the contribution of the subleading term for positive values of $\phi_R$, but enhances this same contribution for negative $\phi_R$. This effect can be traced back to us choosing to set the second bracketed factor in \eqref{eq:V_W_factored} equal to $f$. Had we chosen to set the first factor equal to $f$, the same analysis would result in an approximation $V_{W_2}(\phi_R)$ of $V(\phi_R)$ which is accurate for all $\phi_R<0$, but breaks down for $\phi_R$ positive enough. Since throughout this paper we have worked with the convention that the scalar field rolls along the positive real axis, it makes sense to consider, as we have done, the construction which gives an accurate approximation $V_{W_2}(\phi_R)\approx V(\phi_R)$ for positive $\phi_R$. Nevertheless, by taking $C$ large enough, we can make $V_{W_2}(\phi_R)$ arbitrarily close to $V(\phi_R)$ for $\phi_R$ greater than any real $\phi_R^*$.

\paragraph{Maximum at $\phi_R=0$}

One potential issue with the truncated potential $V_{W_2}(\phi_R)$ is that it is not symmetric with respect to $\phi_R\rightarrow -\phi_R$; see Fig. \ref{fig:V_and_VW2}. Despite this, from \eqref{eq:V_W2} we find 
\begin{align}
    V_{W_2}'(\phi_R) = V'(\phi_R)\left(1+\frac{2V(\phi_R)e^{-2\sqrt{3}\phi_R}}{12 C^2}+\frac{1}{2\sqrt{3}C^2} \int_0^{\phi_R} du \,V(u) e^{-2\sqrt{3}u}\right),
\end{align}
and hence $V_{W_2}'(0) = 0$ if $V'(0) = 0$. Thus, although $V_{W_2}(\phi_R)$ is not symmetric, it does have a local maximum at $\phi_R=0$, allowing for the asymptotically AdS solution $\phi_R=0$ in the Euclidean continuation. Therefore it is not a problem that $V_{W_2}(\phi_R)$ is not symmetric.

\paragraph{Dynamics of a complex $\phi$}

So far we have restricted the complex field $\phi = \phi_R+i \phi_I$ to the real line, i.e. we have set $\phi_I = 0$. This has allowed us to explicitly construct the superpotential $W_2(\phi_R)$, which gives a scalar potential $V_{W_2}(\phi_R)$ that is arbitrarily close to the cosmological potential $V(\phi_R)$. Analytically continuing away from the real axis, we obtain the superpotential $W_2(\phi)$ as a function of complex $\phi$. Notice from \eqref{eq:W2} that if $V(\phi)$---the analytic continuation of the cosmological potential $V(\phi_R)$---is analytic in some simply connected region of the complex plane containing $\phi=0$, then $W_2(\phi)$ is analytic in this same region. The cosmological potentials considered in previous sections were analytic everywhere, and so the corresponding superpotential is also analytic everywhere.

The superpotential $W_2(\phi)$ defines a scalar potential $V_{W_2}(\phi)$ for complex values of $\phi$ via Eq. \eqref{eq:V_W_general},
\begin{align}\label{eq:V_W_general_2}
    V_{W_2}(\phi) = e^{\phi\bar\phi}\left(|W_2'(\phi)+\bar\phi W_2(\phi)|^2 -3|W_2(\phi)|^2\right).
\end{align}
Notice that $V_{W_2}(\phi)$ is not holomorphic even if $W_2(\phi)$ is. In fact, $V_{W_2}(\phi)$ is real for all $\phi$, and so it cannot be holomorphic. Thus instead of thinking of $V_{W_2}(\phi)$ as a function of a single complex variable $\phi$, it makes more sense to think of it as a (real) function of two real variables, $\phi_R$ and $\phi_I$. 

By construction we have ensured that $V_{W_2}(\phi)$ restricted to the real axis is approximately equal to the cosmological potential $V(\phi_R)$. An important question is: if at some initial time the values of $\phi$ and $\dot \phi$ are both real (i.e. $\phi_I=0=\dot\phi_I$), then does $\phi$ remain real (i.e. does $\phi_I$ remain zero) throughout its dynamical evolution? In other words we are asking whether the imaginary component of $\phi$ decouples, leaving only a single real scalar degree of freedom to source the cosmological dynamics.

To answer this, suppose that $\phi$ is spatially homogeneous and consider the equation of motion for the fields $\phi_R$ and $\phi_I$ with potential $V_{W_2}(\phi_I,\phi_R)$,
\begin{align}
    \ddot{\phi}_i + 3 H \dot{\phi}_i + \partial_i V_{W_2} = 0.
\end{align}
Since $W_2(\phi)$ is analytic in the complex plane and real on the real axis, it can be shown that $\overline{W_2(\phi)}=W_2(\bar\phi)$.\footnote{Proof: Since $W_2(\phi)$ is analytic $W_2(\phi)=W_2(\phi_R+i\phi_I) = \sum_{n=0}^{\infty} \frac{1}{n!}W_2^{(n)}(\phi_R)(\phi-\phi_R)^n$. Since $W_2(\phi_R)$ is real $\overline{W_2^{(n)}(\phi_R)} = W_2^{(n)}(\phi_R)$. Hence $\overline{W_2(\phi)}= \sum_{n=0}^{\infty} \frac{1}{n!}W_2^{(n)}(\phi_R)(\bar\phi-\phi_R)^n = W_2(\bar\phi)$.} 
Equation \eqref{eq:V_W_general_2} then implies that $V_{W_2}(\bar\phi) = V_{W_2}(\phi)$ which implies that $\partial V_{W_2}/\partial \phi_I$ is identically zero on the real axis. Therefore if $\phi_I$ and $\dot\phi_I$ vanish at some initial time, they vanish for all times; if $\phi$ starts rolling along the real axis, it stays on the real axis. Hence we are able to construct any effective single field cosmological model, such as those discussed in previous sections, by starting with a supersymmetric Lagrangian.

\paragraph{Stability}

Having shown that $\phi_I=0=\dot\phi_I$ initially implies vanishing $\phi_I$ for all times, i.e. $\phi_I$ decouples from $\phi_R$, it is important to ask whether the dynamics of $\phi_R$ is stable against small perturbations which make $\phi_I$ or $\dot\phi_I$ nonzero. Such perturbations will inevitably spontaneously occur due to quantum fluctuations, so it is crucial to consider their effects.

To determine the effects of such small perturbations away from real values of $\phi$, we must evaluate $\partial^2 V_W/\partial \phi_I^2$ for $\phi_I = 0$. If this is positive, the potential $V_{W_2}(\phi)$ curves upwards as $\phi$ moves away from the real axis, and thus the motion along the real axis is stable to small perturbations. 

Equation \eqref{eq:V_W_general_2} gives $V_{W_2}(\phi)$ for all complex values of $\phi$. Setting $\phi=\phi_R+i\phi_I$ and differentiating, we find
\begin{align}
    \frac{\partial^2V_{W_2}}{\partial \phi_I^2}\Bigg |_{\phi_I=0} 
    =
    8 e^{2\sqrt{3}\phi_R} C^2 + F_0(\phi_R) + \frac{F_2(\phi_R)}{C^2},
\end{align}
where $F_0(\phi)$ and $F_2(\phi)$ are analytic functions (which depend on $V$) and are thus bounded on any compact set $I\subset \mathbb R$. Therefore, since the leading term in this expansion is positive, it is always possible to take $C$ large enough so that $\partial^2 V_{W_2}/\partial \phi_I^2|_{\phi_I=0} >0$ for any $\phi_R\in I$.\footnote{Recall that taking $C$ large is the same limit which ensures $V_{W_2}(\phi_R)\approx V(\phi_R)$ in the first place.} In this case the dynamics of $\phi_R$ along the real axis is guaranteed to be stable under fluctuations in the imaginary directions, and so even at the level of perturbations the field $\phi_I$ decouples from the dynamics of $\phi_R$.\footnote{Note that the dynamics is also stable in the double analytically continued (Lorentzian wormhole) picture \cite{Antonini:2022blk}, because $\partial^2 V_{W_2}/\partial \phi_I^2 >0$ trivially satisfies the BF bound, while $\partial^2 V_{W_2}/\partial \phi_R^2$ satisfies the BF bound by construction.} Thus we truly have a single field cosmological model, starting from a supersymmetric Lagrangian.

\section{Outlook}
\label{sec:outlook}

In this paper, we have argued that accelerated expansion is generic (i.e. arises without fine tuning) in $\Lambda < 0$ cosmologies arising from asymptotically AdS Euclidean wormholes, for simple scalar potentials with a form that is natural in theories with a dual CFT. We found that there exist models of this type which can match $\Lambda$CDM to high accuracy at the level of background geometry, so the framework can give realistic cosmologies. We found that the scalar potential in these examples can arise from a superpotential, so the effective field theory can potentially have a supersymmetric point for $\phi = 0$. There are many directions for further study.

\paragraph{Direct reconstruction of the potential} The class of models considered in this paper make at least two generic predictions: that one should have a dynamical scalar field contributing to the energy density of the universe and that the universe should eventually begin to decelerate.
With this in mind, it is interesting to consider direct measurements of the scale factor from supernova observations and search for evidence of an evolving scalar field. In principle, the scalar potential in a single field model can be reconstructed directly from knowledge of the scale factor $a(t)$ and $\Omega_M$ when radiation can be ignored. The explicit expressions given in equations (\ref{phisol}) and (\ref{V3}) define $V(\phi)$ parametrically. However, $\Omega_M$ (the present day matter contribution to the energy density) is not well constrained by direct observation, and the reconstruction of $V(\phi)$ for given $\Omega_M$ is quite sensitive to small changes in $a(t)$. As a result, there is actually considerable freedom in the form of the potential that can be consistent with data, and thus considerable room for the types of models considered here.\footnote{One example of this flexibility is highlighted in Figure \ref{fig:2022_Nov16_wCDM_pot} which shows in red a potential with significant variation that provides a good fit to the low redshift measurements of the scale factor.} Plausibly, the type of fine-tuning that we required to match $\Lambda$CDM precisely will not be required if the goal is to provide a good fit with direct obserations of the scale factor. We will discuss these issues along with a more complete analysis of the supernova data in a forthcoming paper \cite{dataPaper}.  We note that it is particularly interesting at present to consider models that go beyond $\Lambda$CDM given the Hubble tension \cite{di2021realm}. 

\paragraph{Including fluctuations in a realistic effective field theory model} In this paper, we have focused entirely on the evolution of the homogeneous background. We found that there are effective theories which are compatible both with the cosmological history obtained from $\Lambda$CDM or $w$CDM and with having a stable asymptotically AdS analytic continuation. It is very interesting to understand whether these effective theories are also capable of reproducing the observed spectrum of fluctuations in the CMB. One question we would particularly like to understand is whether primordial inflation is required or if the existence of the analytic continuation is already enough to give a sensible spectrum of fluctuations. 

We also want to understand if the scalar potentials we are considering could have other phenomenological effects. For example, considering fluctuations around a very flat potential raises the prospect of a new long-ranged scalar-mediated force. One can also ask if scalar particles would be produced in any appreciable quantity and, if so, could such particles form a component of dark matter? These are issues that must be addressed to claim a fully realistic model, quite apart from objections to flat potentials arising from Swampland conjectures.

\paragraph{Towards a microscopic construction} Here we have focused exclusively on effective theories, but it remains an important open question to find concrete microscopic realizations of these big bang / big crunch cosmologies with a holographic dual. In the search for such constructions, it makes sense to again back off of demanding a fully realistic model. Instead, we can ask if relatively simple models can be found, perhaps corresponding to a pure radiation scenario or a scenario with a relatively simple scalar potential exhibiting some period of acceleration.

It is also interesting to further explore the issue of supersymmetric effective theories. Here we showed that a superpotential can be constructed to match any potential we devise, but presumably not all such superpotentials arise from a UV-complete and supersymmetric microscopic theory.

\textit{Acknowledgements:} We acknowledge support from the U.S. Department of Energy grant DE-SC0009986 (B.G.S.), the U.S. Department of Energy, Office of Science, Office of Advanced Scientific Computing Research, Accelerated Research for Quantum Computing program ``FAR-QC'' (S.A.), the National Science and Engineering Research Council of Canada (NSERC) and the Simons foundation via a Simons Investigator Award and the ``It From Qubit'' collaboration grant. 

\appendix

\section{Details of the potential examples}

\subsection{Fine-tuning potentials with a flat region}
\label{app:potentials}

In this appendix we explain how the parameters of potentials with a flat region of the form (\ref{eq:flatpot})-(\ref{eq:flatpot2}) must be chosen and fine-tuned to reproduce the $\Lambda$CDM cosmological evolution between the early universe and the present day.

In order to have a phase of accelerated expansion in the cosmological evolution, the positive potential energy at the plateau (which is fixed by observational data to be $A=\Omega_{\Lambda}$) must be dominant over every other term on the right-hand side of the Friedmann equation. This implies that the scalar field's kinetic energy $K_{\phi}$ must be very small when the field reaches the plateau. Evolving backwards in time from the time-symmetric point $\tilde{t}=0$ in the cosmology picture, the scalar field behaves like an anti-damped particle. Therefore, intuitively, an accelerated expansion can take place only if $|B|\ll |A|$ --- where $B$ sets the value of $V(0)$. In this way, for an appropriately fine-tuned potential, the scalar field just barely makes it up to the plateau and the kinetic energy does not become dominant until very close to the Big Bang (or Big Crunch, if we evolve forward in time from the time-symmetric point). 


Next, we must require that the BF bound (\ref{BF}) is satisfied. This can be achieved by choosing the value of $X$ to be large enough that the deep valleys are not too close to the maximum at $\phi=0$, so that $(d^2\tilde{V}/d\phi^2)|_{\phi=0}$ is suppressed. However, $X$ must not be too large, otherwise the cosmological evolution will reach the Big Bang singularity before the scalar field can reach the valleys (and then the plateau): in this case, no accelerated expansion phase occurs. 

Moreover, the (negative) parameter $C$ --- which controls the depth of the valleys --- must be large enough that the scalar field acquires sufficient kinetic energy in its anti-damped motion to reach the positive plateau. But it cannot be too large, or the kinetic energy will be large when the scalar reaches the plateau and no accelerated phase can take place. We remark that no particular fine-tuning has been carried out so far on the parameters: we just identified the general criteria to choose the parameters $B$, $C$ and $X$ in order for an accelerated expansion to be possible and for the BF bound to be satisfied (we remind that $A$ is fixed by data). The specific choice of their values is arbitrary and irrelevant for the sake of reproducing the $\Lambda$CDM evolution.

Finally, we fine-tune the value of $\Delta$\footnote{Note that we could have instead fixed a value of $\Delta$ and fine-tuned the parameter $C$.} to allow the scalar field to reach the plateau with a very small kinetic energy. This yields a phase of accelerated expansion, and will allow us to reproduce the $\Lambda$CDM cosmological evolution between the early universe and the present day. 

Note that, as we evolve backwards in time towards the Big Bang, the scalar field's kinetic energy will eventually become dominant. Our goal is to reproduce as best as possible the results of the $\Lambda$CDM model, in which there is no such term in the Friedmann equation, between the early universe and the present day. More precisely, an important criterion for the feasibility of these models is that we should not have a significant departure from $\Lambda$CDM between the present time and the time of Big Bang Nucleosynthesis (BBN) (which roughly occurs in the range of temperatures $T \sim 10^{9}-10^{10}$ K), otherwise the abundances of various elements could vary appreciably from those experimentally observed. These temperatures correspond to scale factors which are between $10^{-9}$ and $10^{-10}$ times smaller than the current value of the scale factor $a(\tilde{t}_0)=1$. We can therefore ask whether it is possible to tune our model with a rolling scalar such that the scalar field's kinetic energy is a subdominant contribution in the cosmological Friedmann equation for $10^{-10}\lesssim a(\tilde{t})\leq 1$. In other words, we would like to determine whether the kinetic energy dominance can be pushed sufficiently early in time that it would not impact the predictions of BBN. 

By appropriately fine-tuning the parameter $\Delta$ we were able to push the crossover between the early kinetic-energy-dominated era and the radiation-dominated era to $\tilde{t}=\tilde{t}^*$ with $\tilde{t}^*-\tilde{t}_{BB}=\mathcal{O}(10^{-13})$ (where $\tilde{t}_{BB}$ corresponds to the Big Bang) in our numerical solutions, corresponding to $a(\tilde{t}^*)=\mathcal{O}(10^{-7})$. We were not able to reach the desired value $a(\tilde{t}^*)=\mathcal{O}(10^{-10})$ because of numerical precision issues related to the excessively small time steps needed, but there is no a priori obstacle in pushing the kinetic-energy-dominated era arbitrarily close to the cosmological singularity.

By implementing this fine-tuning procedure after choosing the \textit{Planck} 2018 cosmological parameters (\ref{eq:planckdata}) and the initial condition $a_0=2$, we obtained the set of parameters (\ref{eq:numpotparam}) used in our numerical analysis. The same procedure was also used to fine-tune the potential for the solution employing SNIa-derived cosmological parameters depicted in Figure \ref{fig:muflatpot}.

\subsection{Scalar potential for \texorpdfstring{$w$}{}CDM model} \label{app:wcdm}

For completeness, we provide some details related to the analysis in Section \ref{sec:wcdm}. 

To construct a scalar potential $V(\phi)$ agreeing with $V_{w}(\phi)$ in a suitable region, permitting cosmological solutions which satisfy observational constraints from SNe Ia and have asymptotically AdS Euclidean continuations, we proceed as follows:
\begin{enumerate}
    \item Determine the redshift dependence of the scalar field $\phi_{w}(z)$ and potential $V_{w}(z)$ for the given $w$CDM model,\footnote{The Hubble expansion $H(z)$ for the $w$CDM model is given in equation (\ref{eq:H_w}), using the parameters recorded in Section \ref{sec:wcdm}. The redshift dependence of the scalar $\phi(z)$ and potential $V(z)$ may then be computed from $H(z)$ using equation (\ref{eq:Hz_phiz}).} in the region $z \in (z_{\textnormal{min}}, z_{\textnormal{max}})$ where type Ia supernova data have been used to constrain cosmological parameters. We have the freedom to choose the value $\phi_{w}(0) \equiv \phi_{0}$ and the sign of $\phi_{w}'(0)$; we will take both to be positive. Note however that the present scalar kinetic energy is fixed to 
    \begin{equation}
        \frac{1}{2} \dot{\phi}_{w}(z=0)^{2} = \frac{1}{8 \pi G} H_{0} H'(0) - \frac{3}{16 \pi G} H_{0}^{2} \Omega_{M} \equiv K_{0} \: .
    \end{equation}
    \item Evaluate $\phi_{w}(z)$ and $V_{w}(z)$ for a large sample $\{z_{i}\}$ of redshifts in the interval $(z_{\textnormal{min}}, z_{\textnormal{max}})$, to obtain an array of pairs $\{(\phi_{w}^{(i)}, V_{w}^{(i)}) \}$. 
    \item Choose a model $V(\phi; \{p\}, q)$, with parameters $\{p\}$ and $q$, satisfying:
    \begin{itemize}
    \item $V(0; \{p\}, q) < 0$
    \item $- \frac{9}{4} < \frac{V''(0; \{p\}, q)}{|V(0; \{p\}, q)|} < 0$ 
    \item $V(\phi; \{p\}, q)$ behaves as a polynomial for $\phi \gtrsim \phi(z=0)$. 
    \end{itemize}
    In our case, the model has a Gaussian trough between $\phi = 0$ and $\phi = \phi_{0}$, with a height controlled by the parameter $q$.
    \item Apply the ``shooting method" to determine parameters $\{p\}$ and $q$ for which the model has a time-symmetric solution with the correct scalar kinetic energy $K_{0}$ at $z=0$:
    \begin{itemize}
        \item For various choices of $q$, use the array $\{(\phi_{w}^{(i)}, V_{w}^{(i)}) \}$ to determine the best-fit parameters $\{p_{\textnormal{fit}}(q) \}$.
        \item Solve the equations of motion for $a(t), \phi(t)$ for the best-fit models $V(\phi; \{p_{\textnormal{fit}}(q)\}, q)$ with each choice of $q$ to determine for which value $q_{*}$ there exists a future time $t$ for which $\dot{a}(t) = \dot{\phi}(t) = 0$; this determines the model for which the solution has a moment of time symmetry. We use initial conditions $\phi(t_{0}) = \phi_{0}$, $\dot{\phi}(t_{0}) = - \sqrt{2 K_{0}}$, and $a(t_{0}) = 1$.
    \end{itemize}
    \item As a check, evaluate the solution to the model $V(\phi ; \{p_{\textnormal{fit}}(q_{*})\}, q_{*})$ with time-symmetric initial conditions, and verify that the equation of state $w(z)$ behaves appropriately for $z \in (z_{\textnormal{min}}, z_{\textnormal{max}})$, and that the Euclidean continuation has AdS asymptotics.
\end{enumerate}
It will frequently be useful to work with quantities rescaled by the Hubble parameter $H_{0}$, as defined in Section \ref{sec:LambdaCDM}. 

We observe that the procedure outlined above has significant freedom in the choice of model $V(\phi)$; since our goal is merely to demonstrate the existence of a scalar potential satisfying the list of criteria outlined in the main text, it suffices to construct a specific example.

\paragraph{Constructing an example}

We first deduce the redshift dependence of the scalar field $\phi_{w}(z)$ and the rescaled potential $\tilde{V}_{w}(z)$ from the rescaled Hubble expansion $\tilde{H}(z)$ for the $w$CDM model, using
\begin{equation} \label{eq:Hz_phiz}
    \begin{split}
        \phi_{w}(z) & = \phi_{0} + \int_{0}^{z} dz' \: \frac{\sqrt{\frac{1}{4 \pi G}(1+z') \tilde{H}(z') \tilde{H}'(z') - \frac{3}{8 \pi G} (1+z')^{3} \Omega_{M}}}{(1+z') \tilde{H}(z')} \\
        \tilde{V}_{w}(z) & = - \frac{1}{8 \pi G} (1+z) \tilde{H}(z) \tilde{H}'(z) + \frac{3}{8 \pi G} \tilde{H}(z)^{2} - \frac{3}{16 \pi G} (1+z)^{3} \Omega_{M} \: .
    \end{split}
\end{equation}
We also deduce the present time-derivative $\dot{\phi}(\tilde{t}_{0})$, where the overdot denotes a derivative with respect to the rescaled time coordinate; in units $\frac{8 \pi G}{3} = 1$, we find for the $w$CDM model
\begin{equation}
    \dot{\phi}_{w}(\tilde{t}_{0}) = - 0.2628687886 \: .
\end{equation}

Next, we choose as our model function
\begin{equation}
    \begin{split}
        \tilde{V}(\phi; A, B, C, \Delta, X, \mathbf{V}) & = \tilde{V}_{0}(\phi; \ldots) + \tilde{V}_{0}(- \phi; \ldots) \\
        \tilde{V}_{0}(\phi; \ldots) & = \frac{A-B}{2} \textnormal{erf} \left( \frac{\phi - X}{\Delta} \right) + C \exp \left( - \frac{(\phi - X)^{2}}{ \Delta^{2}} \right) \\
        & \qquad + \left( \sum_{k=1}^{4} \frac{V_{k}}{k!} (\phi - X)^{k} \right) \left( 1 + \textnormal{erf} \left( \frac{\phi - X }{\Delta } \right) \right) + \frac{A}{2} \: .
    \end{split}
    \end{equation}
This is similar to the asymptotically flat potential introduced in Section \ref{sec:LambdaCDM} to reproduce the $\Lambda$CDM model, though we have effectively replaced the flat region with a polynomial. 

We will choose to situate the present time at $\phi_{0} = 0.7$, the trough in the potential at roughly $X = 0.5$, the width $\Delta = 0.06$, and to take $B = -0.08$, which fixes the potential at $\phi=0$ to this value; we choose these values for concreteness, though one could proceed analogously for many suitable choices of these quantities. Our choices are motivated by the facts that (1) having the trough join suitably to the polynomial piece of the potential requires $(\phi_{0} - X)$ to be a few multiples of $\Delta$, and (2) $B$ should be sufficiently large in absolute value that we avoid recollapse in the wormhole picture. The parameter $C$, controlling the depth of the trough, is the parameter that we will vary for the shooting method.

With these choices, we proceed to apply the shooting method, solving the equation (\ref{eq:cosmo_evo}) with initial conditions $a(t_{0}) = 0, \phi(t_{0}) = \phi_{0}, \dot{\phi}(t_{0}) = - 0.2628687886$, for various choices of $C$ and best-fit $(A, V_{1}, V_{2}, V_{3}, V_{4})$. 
Our procedure yields, to our working precision, the parameters $C = -7.80623080636$ and
\begin{equation}
\begin{split}
    & A = 1.13258420902349 \: , \quad V_{1} = -4.21938633468535 \: , \quad V_{2} = 50.3190576443637 \: , \\
    & \qquad \qquad \qquad V_{3} = -370.514499837809 \: , \quad V_{4} = 1408.97547041725 \: .
\end{split}
\end{equation}
We plot the corresponding potential $V(\phi)$, as well as $V_{w}(\phi)$, in Figure \ref{fig:2022_Nov16_wCDM_pot} of the main text. We find that the moment of time symmetry occurs at rescaled time $\tilde{t}$ with
\begin{equation}
    \tilde{t} - \tilde{t}_{0} = 0.5712956836 \: ,
\end{equation}
at which time the scale factor and scalar field are given by
\begin{equation} \label{eq:timesymm_a_phi}
    a(\tilde{t}) = 1.41641139203233 \: , \quad \phi(\tilde{t}) = 0.358665688585761 \: .
\end{equation}

To check our results, we can solve both the Lorentzian and Euclidean equations of motion (\ref{eq:cosmo_evo}) and (\ref{eq:eucl_evo}) for the model with scalar potential $V(\phi)$, assuming time-symmetric initial conditions at
\begin{equation}
    \phi = 0.358665688585761 \: ,
\end{equation}
the location suggested by equation (\ref{eq:timesymm_a_phi}). Assuming that the present time $\tilde{t}_{0}$ satisfies $a(\tilde{t}_{0}) = 1$, we can extract the redshift dependence of the equation of state parameter $w(z)$ over the range $z \in (z_{\textnormal{min}}, z_{\textnormal{max}})$ using equation (\ref{eq:w}), and verify that it is approximately constant and equal to the Pantheon+SH0ES value $w = -0.90 \pm 0.14$; we show this in Figure \ref{fig:2022_Nov16_wCDM_sol} in the main text. We can also verify that the Euclidean solution is asymptotically AdS by plotting the rescaled ``Euclidean Hubble expansion" $H(\tau)$ and observing that it approaches a constant value, as also shown in Figure \ref{fig:2022_Nov16_wCDM_sol} in the main text.

\bibliographystyle{JHEP}
\bibliography{refs}

\end{document}